\documentclass[usegraphicx]{mn2e_modmargin}
\usepackage{times,amssymb,amsbsy}
\usepackage[fleqn]{amsmath}

\usepackage{multirow}

\newcommand\bb[1] {   \mbox{\boldmath{$#1$}}  }
\newcommand\del{\bb{\nabla}}
\newcommand\bcdot{\bb{\cdot}}
\newcommand\btimes{\bb{\times}}

\begin{document}
\title[Hysteresis in MRI simulations]{Hysteresis and thermal limit cycles in 
  MRI simulations of accretion discs}

\author [Latter \& Papaloizou]{ H. N. Latter$^{1}$\thanks{E-mail:
    hl278@cam.ac.uk}, J.C.B. Papaloizou$^{1}$\thanks{E-mail: J.C.B.Papaloizou@damtp.cam.ac.uk}\\
$^{1}$ DAMTP, University of Cambridge, CMS, Wilberforce Road, Cambridge, CB3 0WA, U. K.}




\maketitle

\begin{abstract}
The recurrent outbursts that characterise
 low-mass binary systems reflect thermal
 state changes in their associated accretion discs. The observed
 outbursts are connected to 
 the strong variation in disc opacity as hydrogen ionises near 5000 K.
 This physics leads to
  accretion disc models that
 exhibit bistability and
  thermal limit
 cycles, whereby the disc jumps between a family of cool and low accreting states and
a family of hot and efficiently accreting states. Previous models have
parametrised the disc turbulence
via an alpha (or `eddy') viscosity. In this paper we treat the
turbulence more realistically via a suite of numerical
 simulations of the magnetorotational instability (MRI) in
local geometry. Radiative cooling is included via a simple but
physically motivated prescription. We show the existence of
 bistable equilibria and thus the prospect of 
 thermal
limit cycles, and in so doing demonstrate
that MRI-induced turbulence is compatible with
the classical theory. Our simulations also show that the turbulent
stress and pressure perturbations are only weakly dependent on each
other on orbital times; as a
consequence, thermal instability connected to variations in turbulent heating
(as opposed to radiative cooling) is unlikely to operate, in agreement
with previous numerical results. Our work presents a first step towards
 unifying simulations of full MHD turbulence
with the correct thermal and radiative physics of the outbursting
discs associated with
 dwarf novae, low-mass X-ray binaries, and possibly young stellar objects. 
\end{abstract}


\section{Introduction} 
\label{intro}

Recurrent outbursts in accreting systems are commonly
attributed to global instabilities in their associated accretion discs. 
In particular, it is believed that
thermal instability driven by strong variations in the disc's
cooling rate causes the observed
state transitions in dwarf novae (DNe) and low-mass 
X-ray binaries (LMXBs) (Lasota 2001), while the rich array of
accretion variability associated with quasars could be excited by 
thermal instability driven by variations in the 
turbulent heating rate (Shakura and Sunyaev 1976, Abramowicz et al.~1988),
by thermal-viscous instability (Lightman \&
Eardley 1974), or assorted
dynamical instabilities 
(e.g.\ Papaloizou \&  Pringle 1984, Kato 2003, Ferreira \&  Ogilvie 2009).
On the other hand, FU Ori outbursts, characteristic of protostellar discs,
probably result from the interplay of thermal, gravitational, and
magnetorotational  instability (MRI) across the intermittently inert dead zone
(Gammie 1996, Balbus \& Hawley 1998, Armitage et al.~2002, Zhu et al.~2009, 2010).
Exceptions to this class of model include classical novae whose outbursts
can be traced to thermonucleur fusion of accreted material on the
white dwarf surface (Gallagher \&  Starrfield 1978, Shara 1989, Starrfield et al. 2000).

The most developed, and possibly most successful, model of accretion
disc outbursts describes the recurrent eruptions that characterise the light
curves of DNe and LMXBs. 
In this model, the ionisation of hydrogen, occurring at  temperatures of
around 5000 K, 
induces a
significant opacity change in the disc orbiting the primary star
(Faulkner et al.~1983, hereafter FLP83),
which then leads to
thermal instability and hysteresis 
in the disc gas (Pringle 1981).
 The system exhibits a characteristic `S-curve' 
 in the phase plane of surface density $\Sigma$ and central
 temperature $T_c$ (cf. Fig.~1), and  
as a consequence the gas falls into one of two stable
 states:
 an
optically thick hot state, characterized by strong accretion, or an
optically thin cooler 
state, in which accretion is less
efficient (e.g. Meyer \& Meyer-Hofmeister 1981, FLP83).
Outbursts can then
  be modelled via a
 `limit cycle', whereby 
the disc jumps from the low accreting state to the 
high accreting state and then back again as mass builds up and is then evacuated. 
This mechanism, based on local bistable equilibria,
is the foundation for a variety of advanced models, which 
have enjoyed significant successes in reproducing the observed behaviour of
accretion discs in binary systems, even if interesting
discrepancies persist (Lasota 2001).

The classical model of DNe and LMXBs treats the disc as laminar
   and assumes that
 the disc turbulence can be modelled with  the 
 $\alpha$ 
 prescription (Shakura  \&
Sunyaev 1973), whereby
the action of the  turbulent stresses is characterized as a diffusive
process  with an  associated  `eddy' viscosity (Balbus \& Papaloizou
1999). This has been sufficient to sketch out the qualitative
features of  putative outbursts, but such a crude description has
 imposed limitations on the formalism that are now
  impeding  further progress (Lasota 2012).
On the other hand, fully consistent magnetohydrodynamic (MHD)
simulations of disc turbulence generated by the MRI have been
performed for almost 20 years (Hawley et al.~1995, Stone et
al.~1996, Hawley 2000, etc),
 though typically with
simplified thermodynamics (e.g.\ isothermality). 
 It is the task of this paper to begin the process of uniting the classic thermal instability
 models of DNe and LMXBs with full MHD simulations of the MRI, and thus
consistently account for both the turbulence and radiative cooling.
The first step we take is limited to local models,
as even though discs undergo global outbursts, the ability of local annuli
to exhibit hysteresis behaviour is key.
Local studies will permit us to assess if and how the classic model of thermal
instability can  operate in the presence of realistic turbulent
heating. At the same time,
they provide an excellent test of the MRI itself;
 if the MRI is to remain the chief candidate driving disc
 accretion it must fulfil its obligations to classical disc theory.

We undertake unstratified shearing box
 simulations of the MRI 
 that include Ohmic and viscous heating and a radiative cooling prescription
that is able to mimic the transition between the optically thin and thick states
(FLP83). These simulations clearly reproduce S-curves
in the $\Sigma$ and  mean temperature plane, and these constitute
the main result
of our paper. We can thus
animate local thermal limit
cycles 
via a sequence of local box
runs. 
In particular, the simulated turbulent heating is found to be  `well-behaved' and not so
intermittent as to prematurely disrupt the thermal limit cycles
required by the classical theory. 
In addition, simulations with differing net toroidal and vertical
fluxes produce S-curves that exhibit variable mean values
of $\alpha$, suggesting that when global simulations are considered,
and net-flux is no longer conserved locally,
variations in effective  $\alpha$ between the upper 
and lower branches of the S-curve may be produced,
as required by the classical model (e.g.\ Smak 1984a).

Finally, the simulated short-term 
behaviours of the average viscous stress and the disc pressure reveal
only a weak functional dependence,
 as remarked upon
by previous authors (Hirose et al.~2009). This
poor correlation means that proposed thermal instabilities driven by
 a direct heating response to imposed  pressure
perturbations  are  unlikely to function  in
radiation-pressure dominated  accretion discs (e.g.\ Shakura \&  Sunyaev
1976, Abramowicz et al.~1988). This is in marked contrast to the thermal
instabilities implicated in DNe and LMXBs, which are
driven by the cooling response to imposed temperature
perturbations, mediated via opacity variations.
 Note, however, that over \emph{long times} we find that the stress and pressure are correlated.

The plan of the paper is as follows. In the following section we
discuss the thermal instability model for DNe and LMXBs in more detail.
In Section 3 we present the governing equations while outlining the radiative
cooling prescription the simulations adopt.
 In section 4 we give the
numerical details of the simulations.
 Our results are presented 
in Section 5, and we conclude in
Section 6.

\section{Background}
\label{prelim}

We consider close binary systems that consist of one low-mass
lobe-filling star transferring material to a degenerate companion, either a
white dwarf or a black-hole/neutron star. The
characteristic feature of these systems is their recurrent eruptive
activity in the optical or X-ray spectrum, respectively. DNe
undergo outbursts of some 2-5 magnitude which last 2-20 days, with
recurrence intervals   ranging from 10 days to years (see  e.g.\
Warner 1995). In addition, certain sources
exhibit more complex behaviour, such as `superoutbursts' and
`standstills' (whereby the cycle   is interrupted for an indefinite period). 
 LMXBs are poorly constrained relative to DNe because they emit almost exclusively in X-rays,
and so do not enjoy the same observational coverage. 
Even so, it has been established that their outbursts exhibit
luminosity enhancements of several orders of magnitude, while their
spectra progress through a series of canonical X-ray states
(McClintock and Remillard 2006).

It is
generally accepted that outbursts in
these systems take place in the accretion disc  that orbits around the
primary star and are triggered once sufficient mass has built
up in the disc. During an outburst the disc jumps from a cool low-accreting
state to a
hot high-accreting state that deposits this surfeit of material onto the
central object. After this mass has been evacuated the disc returns to
its low state and the cycle repeats. The basic outline  of the
model was  first proposed by Smak (1971) and Osaki (1974); but the
physical basis for instability in the disc
was not understood till much later (FLP83, see also Hoshi 1979).
Researchers now recognise that the cycles are the consequence of
a thermal instability related  to the variation
of opacity with temperature. 

Above roughly $5000$K the collisional 
ionisation of hydrogen commences, and this liberates an increasing number of
free electrons that enhance the gas opacity $\kappa$.
The dependence of opacity on temperature, for a range of densities,
is conveniently illustrated in Fig. 9 of Bell \& Lin (1994).
The figure shows that, after ionisation of hydrogen is complete $ T  \gtrsim
10^4$K, 
we have roughly $\kappa \propto T^{-2.5}$, and thus $\kappa$ 
decreases  with temperature. On the other hand, cold
molecular gas, with $T  \lesssim 2\times 10^3$K, exhibits a $\kappa$ that
also decreases
with
temperature, primarily on account of dust grain destruction. 
But the gas opacity in the
 intermediate temperature regime between these two limits \emph{increases}
rapidly with temperature, because of the flood of free electrons. Here $\kappa \propto T^{10}$, approximately. 
This dramatic tendency for heat to be better trapped as temperature increases
leads immediately to thermal
instability in this regime.

The instability criterion can be formulated quantitatively
  via the following argument. Let us denote by $\epsilon_D$ the local turbulent dissipation rate per
unit area. Then local thermal equilibrium is determined by balancing
this injection rate against the disc's radiative losses: 
\begin{equation}
\epsilon_D = 2\sigma T_e^4,
\end{equation}
where $T_e$ is the effective temperature of the disc's photosphere and
 $\sigma$ is Stefan's constant. If we now denote by $T_c$ the midplane
 temperature of the disc, a local thermal stability
 criterion for this state is simply:
\begin{equation} \label{stability}
 \frac{d\epsilon_D }{d T_c}< 2\frac{d \sigma T_e^4}{d  T_c},
\end{equation}
(Lasota 2001). This expresses the requirement 
that when the midplane temperature $T_c$ is increased,
the cooling response via radiation must outstrip 
the increase in the heating. If this condition is not met, then the
midplane temperature increase will be reinforced and there will be a
thermal runaway.

At high temperatures, well above the ionisation threshold, we find
that 
the disc radiates 
efficiently and that
$d \ln T_e^4/d \ln T_c \approx 7 $. 
The heating rate $\epsilon_D$, on the other hand, increases much more slowly
 ($\propto T_c$ for a laminar $\alpha$ model)
and so thermal stability ensues. 
However,  in the transition regime,
where  $4\times 10^3 < T \lesssim 10^4$ K,
the photospheric temperature $T_e$ varies very weakly with $T_c$ because the opacity 
is increasing rapidly. The cooling rate barely responds to
an increase in central temperature $T_c$, which leads to the trapping of
excess heat and thermal instability.
At the lowest  temperatures, $T \lesssim 4000$K, stability is
recovered because the disc enters an
optically thin regime in which
the rate
of surface radiation increases rapidly with $T_c$.
In summary, the disc is bistable and will tend to
move to one of two thermally stable steady regimes: (a) hot and ionised (hence
efficiently accreting), with $T>10^4$K; or (b) cold and predominantly
neutral
 (less efficiently accreting) with $T<4000$K. In addition, there exists
 an unstable intermediate equilibrium state between these two limits.

\begin{figure}
\includegraphics[width=3in,height=2in,angle=0]{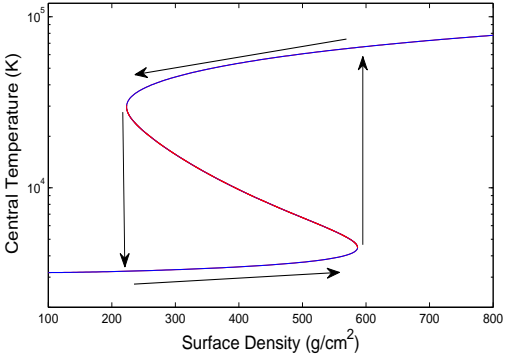}
\caption{Representative plot of an S-curve defining local steady state
thermal equilibria with limit cycle
arrows. This is constructed from  the laminar  $\alpha$  disc
modelling of FLP83. For reference the parameters were $E=3.5$, 
$\lambda=5$, $\mu=1$, and
$\alpha_\text{SS}=0.0275$ (see Section 3 and the Appendix). 
We remark that in this figure, central temperature is plotted as ordinate.
However, if the mass flow rate ${\dot M}$ is used instead,
a curve of the same qualitative form results (FLP83).}\label{fig:1}
\end{figure}

Because for a given surface density $\Sigma$ there can be up to three $T_c$ 
associated with a thermal equilibrium, the family of equilibrium
solutions 
sketches out a characteristic S-curve
in either the $(\Sigma,\,T_c)$ plane  or, alternatively, the $(\Sigma,\,{\dot M})$
  plane, where $\dot{M}$ is mass accretion rate. Note that $\dot{M}$
 can be directly related to $T_c$ and $\Sigma$ (e.g.\ see FLP83 and section \ref{Diagnostics} below).
 In Fig. 1, we plot a representative
 family of such thermal equilibria in  the $(\Sigma,\,T_c)$  plane 
 computed via the techniques of
 FLP83. There is a range of $\Sigma$ for which the
system supports three states, two stable (indicated  with a  blue colour), one
unstable (indicated with a red colour).

In Fig.~1 the trajectory of the
outbursting limit cycle is represented by the black
arrows.
Consider gas physically located at the outermost disc radii and in 
a thermodynamic state associated with the lower branch. 
At this radius, mass steadily accumulates because
accretion onto the primary star cannot match the mass transfer into
the disc from the secondary. Consequently, the surface
density increases and we travel up the lower branch until it ends, at which point the gas
heats up dramatically. Once it settles on the upper branch
accretion is much more efficient and a transition wave propagates through the disc
(Papaloizou et al. 1983, Meyer \& Meyer-Hofmeister 1984, Smak 1984a).
Provided conditions are favourable and the front
propagates inwards, converting the entire disc into an upper state,
 there will be a global
outburst that results in a significant
 mass deposit onto the
primary. Consequently, the
surface density  everywhere decreases until the disc
undergoes a downward transition to the lower branch.

Our discussion of the thermal equilibria and limit cycles
need only consider local models
of an accretion disc. 
However, more advanced DN and LMXB models must incorporate
additional global physics in order to capture the heating and cooling
fronts that mediate branch changes, and ultimately to match
specific observations in detail  (e.g.\ Papaloizou et al.~1983, Smak 1984b, Ichikawa \& Osaki 1992,
 Menou et al.~1999). Even so, all such
  global treatments are founded on the hysteresis behaviour  of the
local model (single annulus)
outlined above. In this paper we concentrate on this fundamental
engine of instability. Our aim is to show that it can function in the presence of
turbulence generated by the MRI. This is intended as a first step in moving
disc outburst modelling away from the heuristic $\alpha$
prescription. Future work can then include the detailed global physics.

\section{Governing equations}
\label{eqns}

 We adopt the local shearing box
 model   (Goldreich \& Lynden-Bell, 1965)
  and a Cartesian coordinate system $(x,y,z)=(x_1,x_2,x_3)$ with
  origin at the centre of the box.
  These coordinates represent the
  radial, azimuthal, and vertical directions respectively. 
 Vertical stratification is neglected.
 The basic equations express the  
 conservation of mass,  momentum, and  energy
and include the induction equation for the magnetic field.
 They incorporate a constant kinematic viscosity $\nu$ and magnetic diffusivity
 $\eta.$ The continuity equation, equation of motion, and induction equations are
given by
\begin{align}
&\frac{\partial \rho}{\partial t} + \del \bcdot (\rho \bb{v})  =  0 \,\label{contg} , \\
&\rho\left( \frac{D\bb{v}}{D t}   + 2 
\bb{\Omega} \times \bb{v}+\del \Phi \right) =\frac{ (\del \btimes
\bb{B})\btimes \bb{B}}{4\pi}
  -\del P +\nabla \bcdot \mathbf{\Pi} , \label{momen} \\
&\frac{\partial \bb{B}}{\partial t}  =  \del \btimes ( \bb{v} \btimes
\bb{B}  - \eta \del \btimes \bb{B} ) \, \label{induct} 
\end{align}
and the first law of thermodynamics is written in the form
\begin{eqnarray}
\rho \frac{D e}{D t} - \frac{P}{\rho}\frac{D \rho}{D t} &=&
\mathbf{\Pi}: \nabla \bb{v} +
 \frac{\eta |\del \btimes \bb{B} |^2}{4\pi}-\Lambda .\label{energycons}
 \end{eqnarray}
Here the convective derivative is defined through
\begin{eqnarray}
\frac{D }{D t} &\equiv &  \frac{\partial }{\partial t}+ \bb{v}\cdot\nabla,
\end{eqnarray}
  $\Omega$ is the Keplerian orbital frequency evaluated at the centre of the box, 
 $\bb{v}$ is the velocity, $\bb{B}$ is the magnetic field
and $e$ is the internal energy per unit mass. 
The tidal potential is $\Phi =-3\Omega^2x^2/2$
and $\Lambda$ is the (radiative) cooling rate per unit volume. The right hand side of
Equation (\ref{energycons})
 includes contributions from  
viscous dissipation, Ohmic heating and radiative losses.
 The components of the
viscous stress tensor $\mathbf{\Pi}$ are 
 given by 
\begin{equation}
\Pi_{ik}=\rho \nu \left( \frac{\partial v_i}{\partial x_k} +
\frac{\partial v_k}{\partial x_i} -\frac{2}{3} \delta_{ik} \del \bcdot \bb{v}
\right) \, ,
\end{equation}
In addition, the magnetic field must satisfy
$\nabla\cdot\bb{B}=0$. Finally,
 we adopt an ideal gas equation of state
\begin{equation} 
 P= \frac{2e}{3}\rho= \rho c_s^2,
\end{equation}
where $c_s^2 = {\cal R}T/\mu$ is the isothermal sound speed, with $T$ the temperature, ${\cal R}$ the gas constant,
and $\mu$ the mean molecular weight. 

\subsection{Radiative cooling model}

In order to proceed, we need to realistically account for
the radiative cooling of the gas, via the term
$\Lambda$. The shearing box by definition is located far from the
disc's upper and lower surfaces with  periodic
boundary conditions applied in the vertical direction. Nevertheless, we can approximate
 the box's radiative losses in a physically meaningful way via
 a prescription
outlined in FLP83.

Our model assumes that the cooling $\Lambda$ is a function only
of time and horizontal position $(x,\,y)$. It is taken to be
\begin{equation} \label{cooling}
\Lambda = 2 \sigma T_e^4/H_0,
\end{equation} 
where $T_e$ is  interpreted as the effective temperature
at the upper and lower vertical boundary of the disc. Here $H_0$ is a
reference scale height. 
The main task is to derive how the surface temperature $T_e$ varies in
response to changes in the midplane temperature  $T_c$.

We envisage that $T_e$ takes a different form in
 three distinct physical disc regimes\footnote{FLP83 originally introduced 4
   regimes, with an extra regime corresponding to the bottom right
   `corner' of the S-curve where the gas may be optically thick. It is
 in fact unnecessary to treat this regime separately, as it is covered
by regimes 2 and 3.}. 

\begin{enumerate}
\item The first regime corresponds to the hot optically thick conditions
associated with the disc's high accreting state. In such a state, the
disc's photospheric temperature $T_e$ is marginally above the value appropriate
for the ionisation
threshold for hydrogen ($\sim 5000$ K) and so the entire disc is optically thick. 
The   classical relationship $T_e^4 =
(4/3)(T_c^4/\tau_c)$, where $\tau_c$ is midplane optical depth,
can be used to relate $T_e$ and $T_c.$ As most of the optical depth
comes from the regions close to the midplane, $\tau_c$
can be evaluated using the midplane opacity.

\item The second regime corresponds to an intermediate `hybrid' state
in which the midplane is hot and partially ionised but the
 surface layers have become sufficiently cool for
hydrogen  ionisation to fall off sharply. As a consequence, 
the opacity drops near the photosphere
with drastic consequences for the disc's vertical temperature structure, as
in cool stars (Hayashi \& Hoshi 1961). 
The classical
dependence of $T_e$ on $T_c$ breaks down, with the connection between
the two becoming quite weak. Instead, we must determine $T_e$ from the outer
 boundary condition $\tau_e=1$, where $\tau_e$ is photospheric
optical depth. 

\item The third regime corresponds to the cool optically thin regime in
which the entire disc, including the midplane, lies marginally below
the ionisation threshold.  In this regime,  consideration of direct cooling 
  then yields  a simple
approximate relationship between $T_c$ and $T_e$.  

\end{enumerate}

 In
 the Appendix we
 consider each regime in turn and obtain three functional forms  relating  $T_e$ 
 to  $T_c$ and $\Sigma$. Our final expression for $T_e$
 is an
 interpolation formula, Eq.~ \eqref{interp}, connecting these different regimes. 
  In summary, the prescription
for finding $T_e$ is given by
\begin{align} \label{Te}
T_e = \begin{cases} \left(4/3\tau_c\right)^{1/4}T_c, &\text{in Regime 1},\\
    \left(10^{36}\,E \rho_c^{-1/3}/\Sigma\right)^{1/10},  &\text{in Regime 2},\\
   \left(2\lambda \tau_c\right)^{1/4}T_c,  &\text{in Regime 3}, \end{cases} 
\end{align}
where $E$ and $\lambda$ are two dimensionless constants, and the
midplane optical depth is
$\tau_c=\kappa_c\Sigma$ with $\kappa_c$ being the opacity evaluated at
the midplane density $\rho_c$
and temperature $T_c$. All dimensional quantities are in cgs units.
 As in FLP83 the opacity in regime 1 is taken to be
\begin{equation}
\kappa = 1.5\times 10^{20}\rho T^{-2.5}
\end{equation}
 and in regime 3 it is taken to be
\begin{equation}
\kappa = 10^{-36}\rho^{1/3} T^{10}.
\end{equation} 
In addition, 
\begin{equation}
\rho_c = \frac{\Sigma}{2H}=\frac{\Sigma\,\Omega\,}{2}\left(\frac{\mu}{\mathcal{R}T}\right)^{1/2}.
\end{equation}
Finally, we specify a constant $\mu$ for all regimes.
This of course neglects partial ionization, but should not interfere
especially in our qualitative results.

In FLP83, thermal equilibrium solutions are computed
by matching the cooling rate, computed according to the
above procedure, with the turbulent alpha heating,
for which the volume heating rate is
\begin{equation} 
\epsilon_D=3\alpha_\text{SS} \rho {\cal R}T \Omega/(2\mu),\label{heating}
\end{equation}
where the Shakura-Sunyaev $\alpha_\text{SS}$ parameter is a dimensionless constant
less than one (cf.\ Eq.~\eqref{SSalph}).
These solutions may be viewed as relations
 between $\Sigma$ and $T_c$, though they
 are more commonly re-expressed as relations between $\Sigma$ and $T_e.$
Some of these S-curves are illustrated in Figs \ref{fig:1} and 4-6.

\section{Numerical set-up}

We  solve
equations \eqref{contg}-\eqref{energycons} with 
 a parallel version of ZEUS
(Stone \&  Norman 1992a, 1992b).  This
uses  a finite difference scheme to obtain the spatial derivatives, 
is first order explicit in
time, and employs constrained transport to ensure
 $\bb{B}$ remains solenoidal. Our version of ZEUS
 has been altered according to improvements outlined by
Lesaffre \&  Balbus (2007), Silvers (2008), and Lesaffre et
al.~(2009).
These simulations are conducted in a shearing box with  dimensions
$(H_0,4 H_0,H_0)$.
As a check on the robustness of the main results
and to verify their independence on the numerical implementation,
 some simulations were also
performed with NIRVANA (Ziegler 1999,   Papaloizou \& Nelson  2003)
adopting a shearing box with dimensions
$(L_x,L_y,L_z)=(H_0,\pi H_0,H_0).$

The reference scale height is given by
\begin{equation}
H_0= \frac{c_{s0}}{\Omega}=\left(\frac{\mathcal{R} T_0}{\mu\Omega^2}\right)^{1/2}
\end{equation}
and thus corresponds to some prescribed isothermal sound speed
$c_{s0}$ or, equivalently, temperature $T_0$.
Note that as the simulations considered here are not isothermal,
$c_{s0}$ need not necessarily correspond to the
actual sound speed at any location or any time. In practice, however,
 it often approximates a volume average of the initial average sound speed
(for  ZEUS runs) or
 the sound speed
under quasi-steady conditions (NIRVANA runs). 

The box is periodic in the
$x$ and $z$ direction and shearing periodic in the azimuthal $y$
direction (see Hawley et al.~1995). Space is scaled by $H_0$ and
time is measured in units of  $\Omega^{-1}$. Density is measured in
terms of the average mass density in the box $\rho_0$, and magnetic
field by the background field strength $B_0$ (see below). Finally,
temperature is scaled by $T_0$.
  
The NIRVANA simulations employ a
resolution of
$(N_x,N_y,N_z)=(128,200,128)$, where $N_i$ is the number of grid cells
in the $i$'th direction. This resolution level was adopted in Fromang et
al.~(2007) and Fromang (2010)  and was there found to be converged. The ZEUS simulations, on the
other hand, adopt  a
 resolution of $(128,512,128)$. Though more
demanding,  because of the higher resolution in the direction of shear,
 the ZEUS simulations can take
advantage of the numerical benefits of an isotropic grid
(Lesaffre \&  Balbus 2007). Additional  runs were performed with the  resolution
 reduced by a factor of two for both codes and these  indicated
  no significant differences from the main results (see Section 5.4).

 In order to study simulations with a range of strengths
of turbulent activity,
 three different magnetic field
configurations (and boundary conditions) were implemented:
fields with zero net-flux (or zero vertical-flux), fields with net vertical-flux,
 and fields with net toroidal-flux (see
Hawley et al.~1995). 
In the net field cases, we must stipulate the
initial plasma beta of a run:
\begin{equation}
\beta =  \frac{8\pi P_0}{B_0^2},
\end{equation}
where $B_0$ is the strength of the  mean field penetrating the box
 and $P_0= \rho_0 H_0^2\Omega^2$ is the
initial pressure.
 As $\beta$ decreases, simulations become more active and
display larger effective values of $\alpha$.
 Furthermore, the cases  with net vertical and 
with net toroidal flux produce turbulence manifesting similar
mean values of $\alpha$ but with different
characteristics  (Hawley et al.~1995).
This can give an indication of to what extent specification of
the mean value of $\alpha$ is sufficient to characterize the thermal equilibria.

 In the zero flux case, we set the initial vertical field to
be of the form  $B_0 \sin(x/H_0)$. 
 However, provided
$B_0$ is large enough for the MRI to be resolved,
the system attains a state after some 10 orbits 
 that is independent of the initial condition (Balbus \& Hawley 1998).

 For the diffusivities,  we  adopted parameters similar to those
 in Fromang et al.~(2007)  which compared
 results for zero flux simulations obtained from three 
 codes, ZEUS, NIRVANA and the PENCIL code.
 These were found to be consistent and, in subsequent work
of Fromang (2010), converged. We remark that Fromang et al.~(2007)
showed that the level of MRI turbulence critically
depends on the magnetic Prandtl number $\mathrm{Pm}=\nu/\eta$, i.e.\ the ratio
of kinematic viscosity to magnetic diffusivity. As a consequence,  we
set $\mathrm{Pm}=4$ to ensure sustained activity, and take
\begin{equation}
\nu=32\times{}10^{-5}H_0^2\Omega \quad \text{and} \quad \eta=8\times 10^{-5}H_0^2\Omega.
\end{equation}

Finally, in the case of NIRVANA, the initial data was prepared from similar 
isothermal simulations
to those carried out in Fromang et al.~(2007). Heating and cooling is
then `switched on' at a given time
after a quasi-steady turbulent state is attained. The ZEUS runs begin
from large amplitude perturbations which blend a large number of MRI
shearing waves, as calculated from the linear theory (Balbus \&
Hawley 1992). In this way a saturated turbulent state can be achieved
relatively quickly and without an initial disruptive transient
(Lesaffre et al.~2009).

\subsection{Numerical implementation of the radiative cooling model}
In this paper we are interested in thermal equilibria
generated by models in which 
heating is provided directly through MRI turbulence.
Accordingly, the cooling prescription must be implemented
as an algorithm within our simulations.
 The values for $T_c$ that are input into the function $\Lambda$
  (see equations (\ref{cooling}) and (\ref{cool}))  is taken to be
the vertically averaged temperature in the box, 
\begin{equation}
T_c(x,y,t) = \frac{1}{L_z} \int_0^{L_z} T(x,y,z,t)\,dz,
\end{equation}
and thus
 $\Lambda$ is a function only of $x$, $y$, and time. Consequently, 
the cooling rate is the same for every $z$ at a
given horizontal location $(x,y)$. This is a natural consequence of using a
one-zone cooling model of the type we have adopted. 

As mentioned earlier, the governing
equations that we numerically solve are scaled according to fiducial
values for time, space, density, etc. However, when we evaluate the
cooling function $\Lambda$, we must convert back to physical dimensions. This
means we need to specify explicit values for $\Omega$, $T_0$, and
$\Sigma$. All calculations were carried out adopting an angular velocity $\Omega$
corresponding to a Keplerian circular orbit of radius $3\times
10^{10}$ cm
around a star of mass $1M_{\odot}$. This choice 
corresponds to the outer regions of an accretion disc around a white dwarf
and yield $\Omega=2.22\times 10^{-3}$ s$^{-1}$ (FLP83). We explored,
however, a range of values for 
the temperature scale $T_0$ and $\Sigma$, over separate simulations, in
order to sketch out the various thermal equilibria in the phase space 
of these variables. The values were
chosen so that the simulation begins near an estimated point on an
S-curve calculated from an alpha-disc model. Once $\Lambda$ is calculated in dimensional units it is
rescaled to  simulation units and fed into the energy equation
\eqref{energycons}. 
 In most cases, simulations remain in the same phase space  neighbourhood.
However departures can occur in cases where there is a thermal
runaway (i.e. no nearby equilibrium). In  some  cases the box dimensions
become mismatched to the physical conditions, at which point  the simulation was terminated.

While $T_0$ and $\Sigma$ varied considerably, we employed two
parameter sets for $(\mu,E,\lambda)$. The first set, corresponds to:
 $\mu=1$, $E=3.5$, $\lambda=5$. We refer
to these as `Set 1'. The second set
 was that
adopted by FLP83: $\mu=0.5$, $E=5.66$, and $\lambda=1$. We refer to
these as `Set 2'. We choose two sets to extend the range of possible
disc states and also display the insensivity of our qualitative results
to these parameters.

\subsection{Diagnostics}\label{Diagnostics}

 Of central importance in our simulations are the turbulent stresses
which, via the transport of angular momentum, facilitate mass
accretion and localised heating (see Balbus \& Papaloizou 1999).
We measure the stresses via the turbulent alpha:
\begin{equation}
\alpha= \frac{\langle T_{xy} \rangle}{\langle P\rangle}
\equiv  \frac{\langle\rho v_x v_y-
B_{x} B_{y}/(4\pi)\rangle}{\langle P\rangle} 
\end{equation} 
where $T_{xy}$ is the $xy$ component of the total turbulent stress tensor
and the angle brackets indicate a
box average. This is a quantity that fluctuates in time, and thus must
be distinguished from the (constant) Shakura-Sunyaev alpha parameter $\alpha_\text{SS}$ that is used
in laminar disc modelling, defined through
\begin{equation} \label{SSalph}
\rho\nu = \frac{2}{3}\alpha_\text{SS}\,\frac{P}{\Omega}.
\end{equation}
The two quantities should coincide once the turbulent alpha is averaged over
long times. 

The turbulent stresses $T_{xy}$ extract energy from the
orbital shear and, once this energy travels down a turbulent cascade,
it is thermalised by Ohmic and viscous dissipation. This turbulent
heating is captured directly in our model via the first two terms on
the right side of \eqref{energycons}. In a thermal quasi-steady state
this heating rate $\epsilon_D$ must balance the radiative cooling rate
$\Lambda$. If this is not achieved the temperature of the system $T$
will evolve until a thermal balance is
obtained.

In a steady state laminar disc the accretion rate ${\dot M}$
is given by 
 \begin{equation} 
{\dot M}= 3\pi{\bar \nu}\Sigma,
\end {equation}
 where ${\bar \nu}$
is the vertically averaged value of $\nu$ weighted by density.
 In a turbulent disc, the equivalent relation  
employing local averaged quantities is
 \begin{equation} 
{\dot M}= \frac{2\pi}{\Omega} \int^{\infty}_{-\infty}\langle T_{r\phi}\rangle
  dz \approx 2\pi H_0\,\frac {\langle T_{xy}\rangle}{\Omega},
\label{pp1}
\end {equation}
and, like $\alpha$, is a time varying quantity.

\begin{figure} 
\centering
\scalebox{0.35}{
\includegraphics{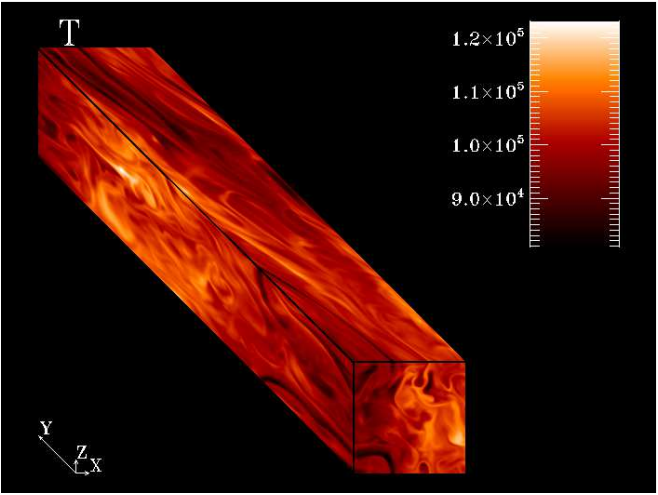}}
\caption{ A screen shot of the temperature in Kelvin
  during a saturated equilibrium
 state. Parameter Set 1 is used with $\Sigma=1000$ g/cm$^2$ with net toroidal
 field. 
The upper branch
 solution has been achieved.}\label{fig:3}
\end{figure}

\begin{figure*}
\centering
\includegraphics[width=7.5in,height=5.5in,angle=0]{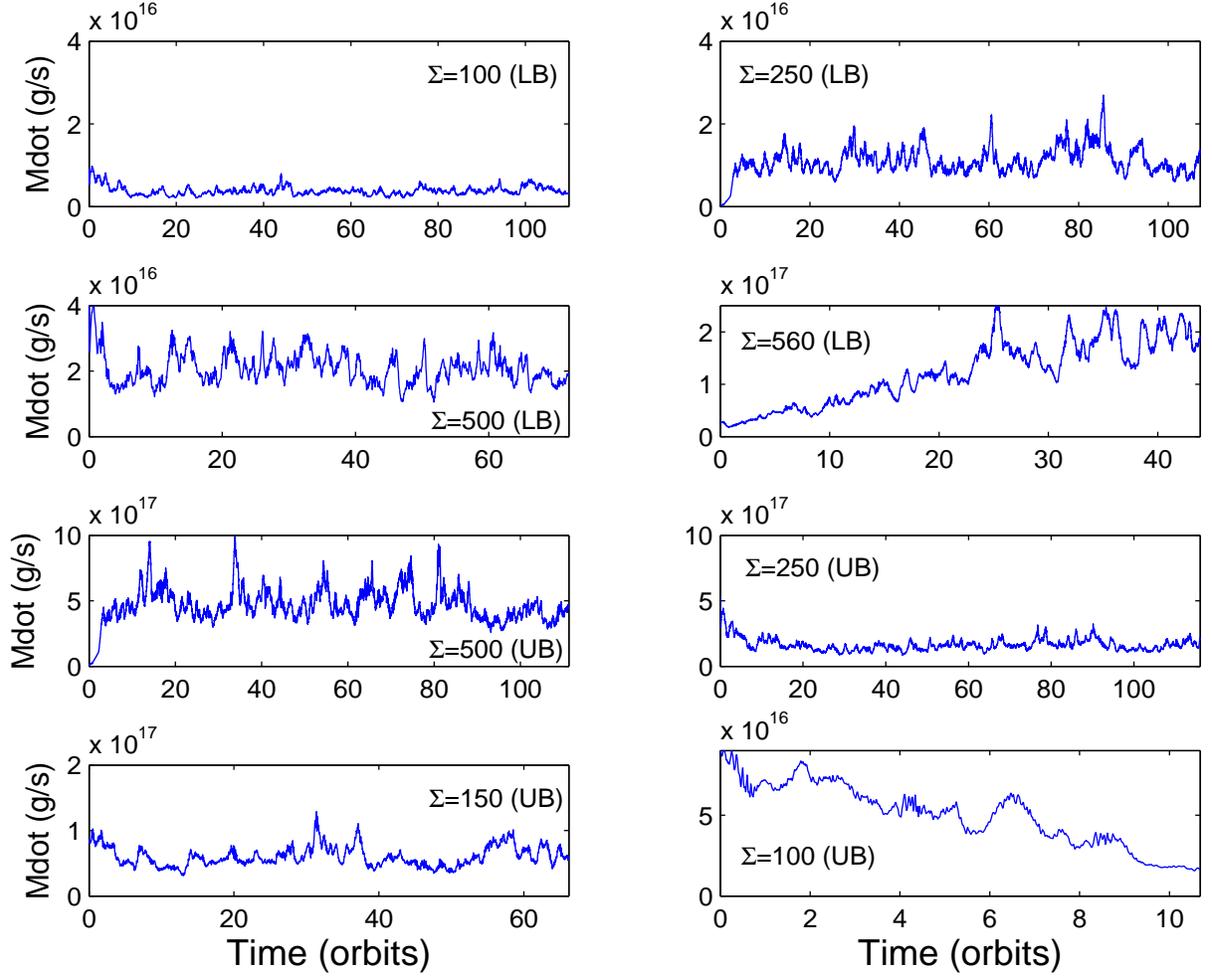}
\caption{ The time evolution of the mass accretion rate
  $\dot{M}$ in eight simulations for the case of net toroidal-flux and parameter
 set 1. Here $\beta=100$. 
 Each simulation is distinguished by its (conserved) surface
 density $\Sigma$ in cgs units. 
In a sequence running from left to right and top to bottom,
the conserved surface density moves first through increasing values 
for which there is a mean steady state corresponding to the lower 
branch (LB) of the S-curve plotted in Fig.~\ref{fig:2rr}. The simulation that is 
second from the top and on the right then moves  to the upper branch (UB)
as there is no available  mean steady state on the lower branch.
The sequence resumes with states of decreasing surface
density tracing out the hot high-accreting solution on the upper branch. 
The final bottom right simulation then
transitions to the lower branch, there being no
 steady state available on the upper branch for its low $\Sigma$. }\label{fig:1rr}
\end{figure*}

\section{Numerical results}
\label{results}

We undertake  numerical simulations of MRI-induced
turbulence
for two sets of cooling parameters. For each
parameter set we investigate three different magnetic
configurations: (a) zero-net flux, (b) net-vertical flux, and (c)
net-toroidal flux. And for each configuration we   conduct a suite of
separate simulations with various $(\Sigma,\,T_0)$ in order to sketch
 S-curves in each of the 6 cases.

 In the net-vertical flux runs, the
initial plasma beta is $\beta=10^4$. This large value is taken in order
to minimise recurrent channel flows --- an artefact of small boxes --- that
may skew our equilibrium results (Hawley et al.~1995, Sano  \&
Inutsuka 2001, Latter et al.~2009). For the net-toroidal flux,
stronger fields were employed and the initial beta was either
$\beta=50$  or $100.$

Once heating and cooling were activated, each simulation
 was run until the system had relaxed towards  a turbulent thermal
 equilibrium, if one was nearby, or had monotonically heated up or
 cooled down so as to change solution branches (cf.\ the vertical
 arrows in Fig.~1). In the former
 cases, the relaxation time depended on the heating rate and hence the
magnitude of the turbulent stresses.
Because of the small values of $\alpha$ in the zero flux runs,
this process could take some time and these runs were typically  continued for
between 150 and 250 orbits. The net toroidal-flux runs exhibited larger
$\alpha$  and a corresponding shorter heating time scale.
They required 30 - 100 orbits to reach a
quasi-steady states with meaningful statistics. 
Fig.~\ref{fig:3} shows
  a screen shot of the turbulent vertically-averaged temperature of
 a representative run 
 once a thermal equilibrium has been achieved. Here a net-toroidal
 flux simulation has been selected
 and the upper branch achieved. 

\begin{figure}
\centering
\includegraphics[width=3.5in,height=2.5in,angle=0]{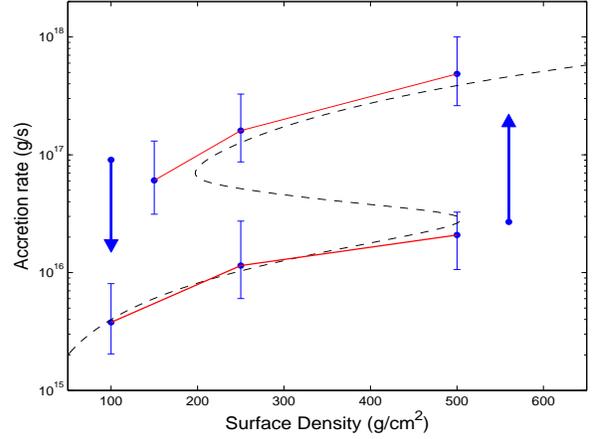}
\caption{  This figure
 summarises the simulations of Fig.~\ref{fig:1rr} by plotting either
 their time averaged $\dot{M}$ in the $(\Sigma,\,\dot{M})$ plane, if a
 thermal equilibrium is achieved, or by showing the initial and end
 $\dot{M}$ if the simulation exhibits a thermal runaway. Equilibria
 are plotted as dots with the error bars indicating the range of
 fluctuations. Runaways are plotted as arrows. Note that despite the
 range of the fluctuations, equilibria (dots) never flipped between
 branches. Only simulations begun near branch edges transitioned. 
 The analytic S-curve (dashed), provided for reference,
is obtained from following the
procedure of FLP83 for $\alpha_\text{SS} = 0.034.$   }\label{fig:2rr}
\end{figure}

In the next subsection we present a collection of 8 runs
associated with parameter set 1 and a net-toroidal flux.
We show their thermal solution trajectories as functions of time and
how these trajectories change as we vary
the surface density $\Sigma$. In
this way we can trace out individual thermal states along a limit
cycle in the $(\Sigma,\,\dot{M})$ plane. Our full set of numerical
results are then displayed with a more detailed discussion
 in the following subsection, where S-curves
are sketched out for all magnetic configurations and parameter
sets. These, however, are presented in terms of $\Sigma$ and 
box-averaged temperature. Once this is done we discuss the weak dependence between
the turbulent stresses and the heating, and finally issues of numerical
convergence.

\subsection{Simulation tracks in the $(\Sigma,\,{\dot M})$ plane}

 In 
laminar disc modelling it is common to describe
 local S-curves and limit cycles within the
 phase plane of $\Sigma$ and $\dot{M}$ because it is the accretion rate
 $\dot{M}$ that dominates the state of the disc.
 Accordingly, we begin by discussing a subset of our simulation results
 in terms of the evolution of $\dot{M}$.
 
In Fig.\ \ref{fig:1rr} we plot the evolution tracks of eight
simulations with net-toroidal flux, with $\beta=100$, and parameter set 1. The
simulations differ in their (conserved) surface density $\Sigma$ and initial
temperature $T_0$.
The simulations
achieve either a thermal quasi-steady equilibrium near their initial state, 
in which $\dot{M}$ fluctuates
about a well-defined mean value,
or they catastrophically heat
up or cool down from their initial state, with an accompanying increase/decrease
in $\dot{M}$. These results are
summarised by Fig.~\ref{fig:2rr} in the
$(\Sigma,\,\langle\dot{M}\rangle)$ plane, 
where the $\langle \dot{M} \rangle$ of each simulation is
computed from a time average over the course of the run. Runs that
achieved a thermal steady state near their initial condition are
represented as dots with error bars showing the range of the
fluctuations in $\dot{M}$. Runs that evolved significantly away from
their initial condition are represented by arrows, the base of which
denote their starting point and the tip of which their end point
when the simulation was terminated. 

As we progress from left to right and top to bottom
 through the eight panels in Fig.~\ref{fig:1rr}
 we track the S-curve in Fig.~\ref{fig:2rr} starting at the bottom
 left corner. First we travel
  towards the right along the lower branch, then up to the higher branch,
 then to the left along the upper branch, and finally back down to the
 lower branch. 
 For the first three
 simulations/panels,
  the surface density
 $\Sigma$ moves through increasing values, each of which permits a
  cool low-accreting equilibrium. But the fourth simulation rapidly
  migrates from the vicinity of the lower branch: no cool 
low-accreting steady state is available, because it possesses a $\Sigma$ that is
  too large. It instead heats up and approaches an upper
  branch of hot solutions characterised by efficient accretion.
   The sequence
  then resumes with a procession of hot equilibria with decreasing
  $\Sigma$. The eighth simulation
transitions from the vicinity of the upper branch to the lower branch
 as there is no hot high-accreting state available that corrresponds to its low
 surface density.

Though a \emph{single} run in local geometry cannot
 describe a complete limit cycle from beginning to end, our
 sequence of \emph{multiple} runs can describe the constituent parts
 of a cycle, state by state. This is permitted because of the
 separation of time-scales between the fast turbulent/thermo dynamics
 that determine the equilibria
 ($\gtrsim 1/\Omega \sim 10^3$ s) and the slow dynamics of the cycle, which
 is governed by the accretion rate ($\gtrsim 1/(\alpha\Omega) \sim 10^5$ s).

 Despite the
large 
range of the $\dot{M}$ fluctuations witnessed in Fig.~\ref{fig:2rr}
there is never any danger that the equilibrium states spontaneously
undergo transitions. States on the lower and upper branches are robust
and distinct, with their stability tied to (the smaller) variations in $T$
rather than $\dot{M}$. This point is is more apparent when the
S-curves are plotted in the $(\Sigma,\,T)$ plane (see next subsection).

Finally, we have superimposed on Fig.~\ref{fig:2rr} an analytic S-curve
computed using the formalism of FLP83 with a Shakura-Sunyaev 
alpha of $\alpha_\text{SS}=0.034$.
The location of the two numerical branches and the stable analytic
branches is reasonably consistent. However, the numerical branches
 often seem more extended to the right.

\begin{figure*}
\centering
\includegraphics[width=3in,height=3in,angle=0]{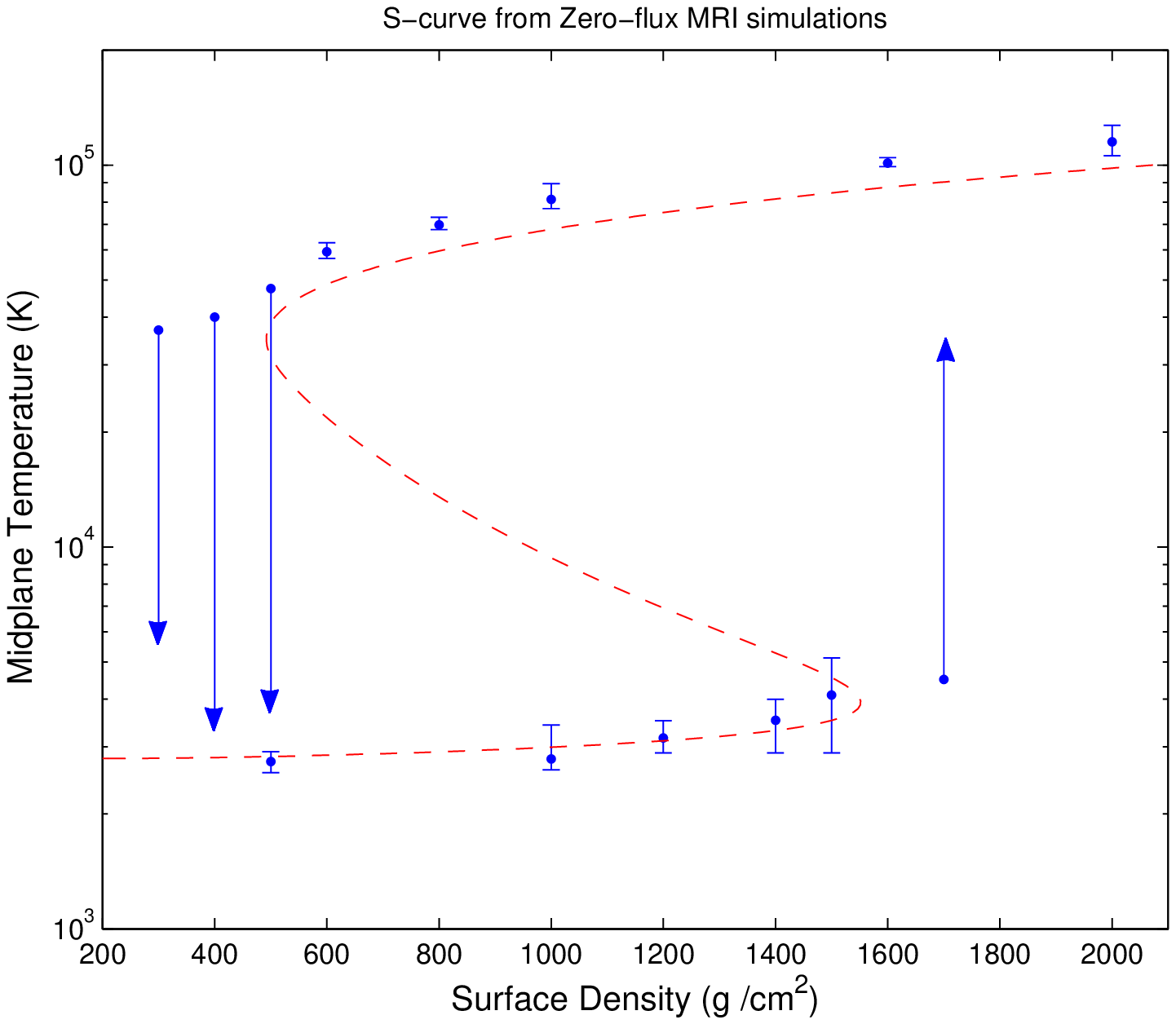}
\includegraphics[width=3in,height=3in,angle=0]
{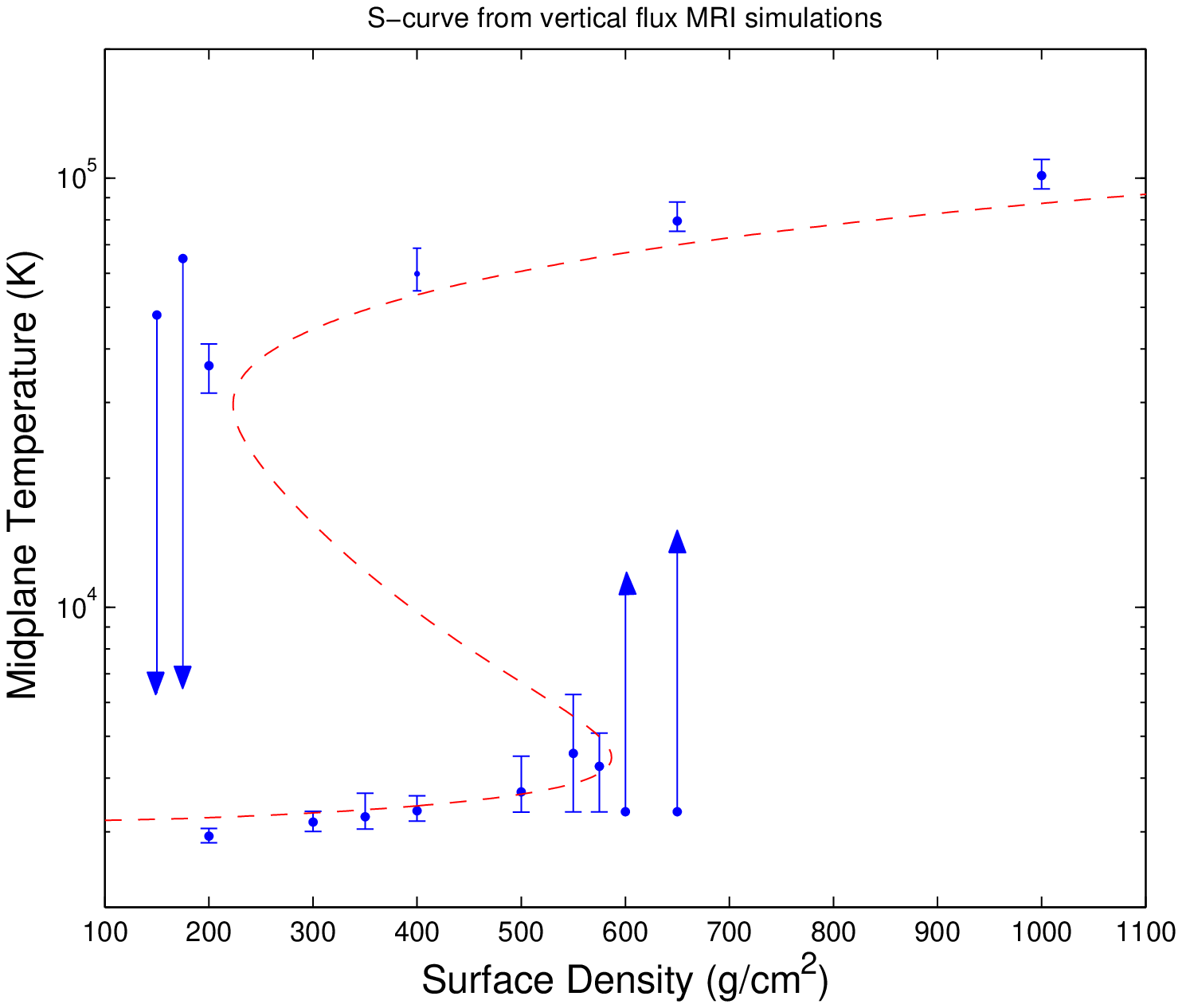}
\includegraphics[width=3in,height=3in,angle=0]
{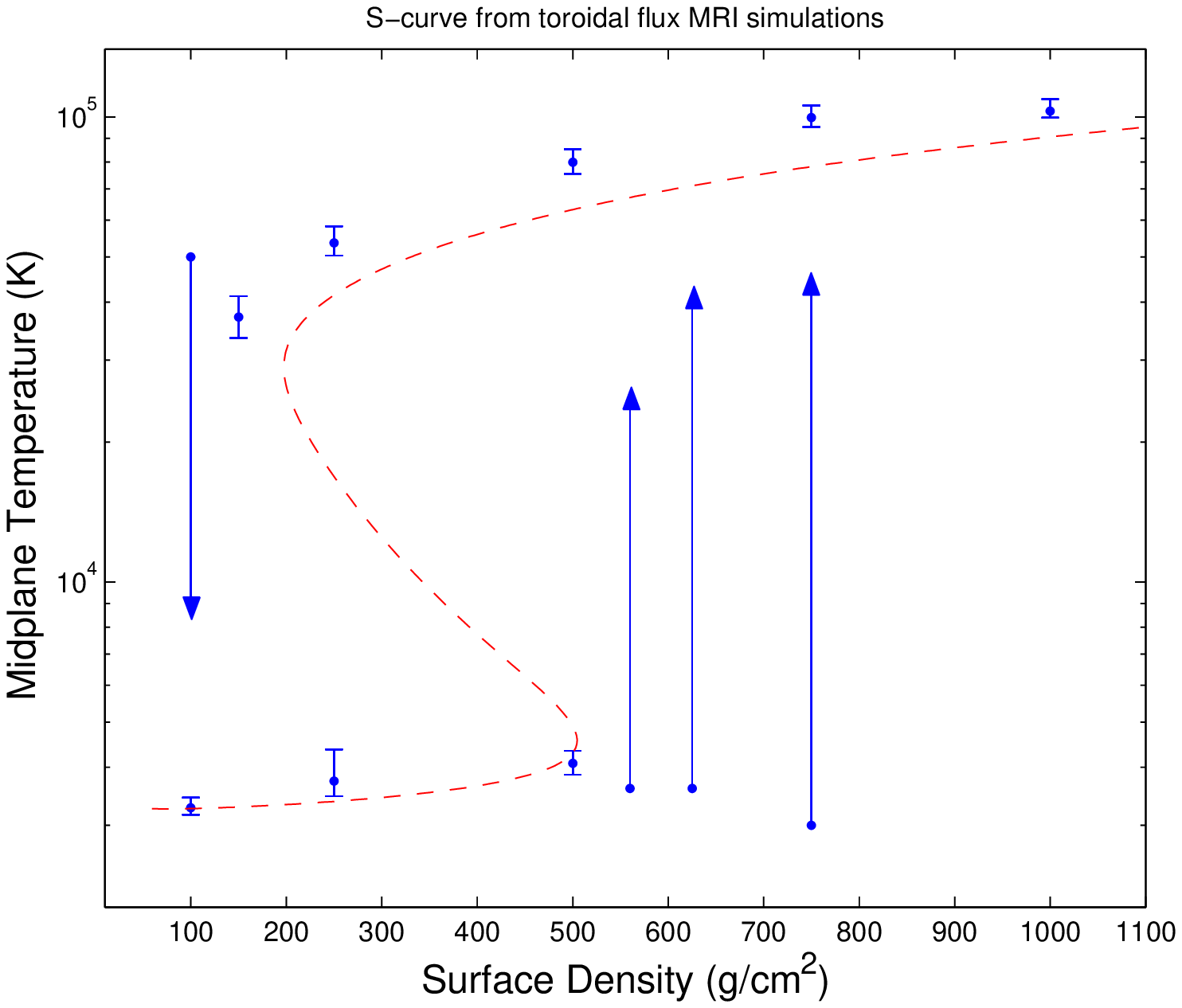}
\caption{S-curves computed with ZEUS for parameter set 1:
  $(\mu,E,\lambda)=(1,3.5,5)$ and for the
 three magnetic configurations. The first panel corresponds to
 zero net-flux, the second panel to net vertical-flux
 (initial $\beta=10^4$), and the third panel to
 a net toroidal-flux (initial $\beta=100$).
Simulation results are indicated on the $(\Sigma, T_c)$
plane, with $T_c$, in K, evaluated as the volume average of the temperature, $T,$  over the box.
$\Sigma$ is expressed in cgs units.
Runs that attained an approximate quasi-steady state are indicated
by a dot, corresponding to the time average  mean volume averaged $T$, and with
 errorbars, indicating the maximum and minimum volume averaged  $T$ once the steady
 state is achieved.
Cases indicated with arrows pointing upward/downward  underwent a heating/cooling
instability, with the volume averaged $T$ on average showing a monotonic
increase/decrease with time.
 Notation is otherwise the same as in
  Fig.~\ref{fig:2rr}.
The dashed curves represent equilibrium S-curves
calculated using the formalism of FLP83 for  $\alpha_\text{SS}=
0.007,\, 0.0275,\, 0.034$ for the zero flux, net-vertical flux, and
toroidal flux cases respectively.
 }\label{fig:5}
\end{figure*}

\begin{figure*}
\centering
\includegraphics[width=3.0in,height=3in,angle=0]{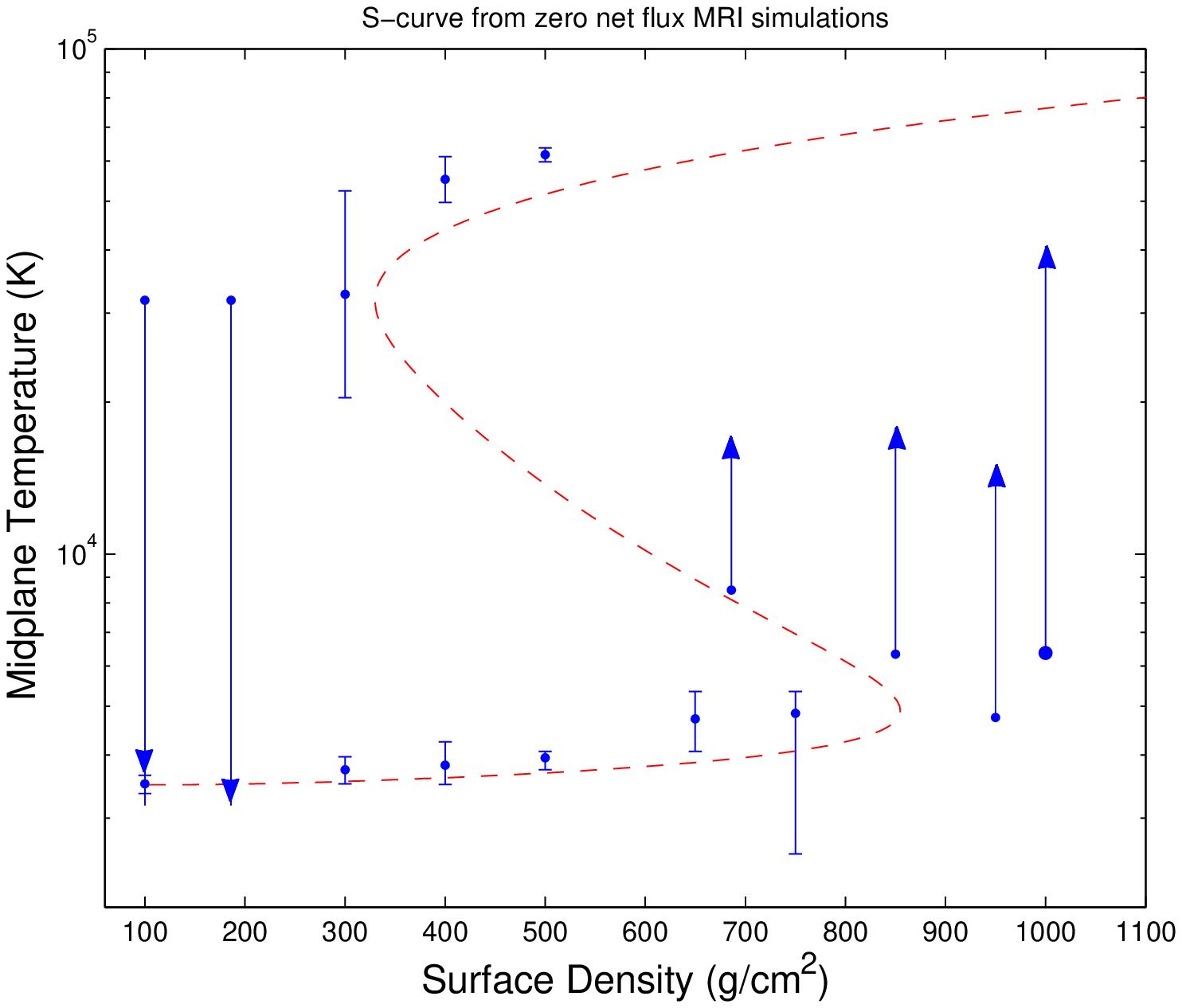}
\includegraphics[width=3.0in,height=3in,angle=0]{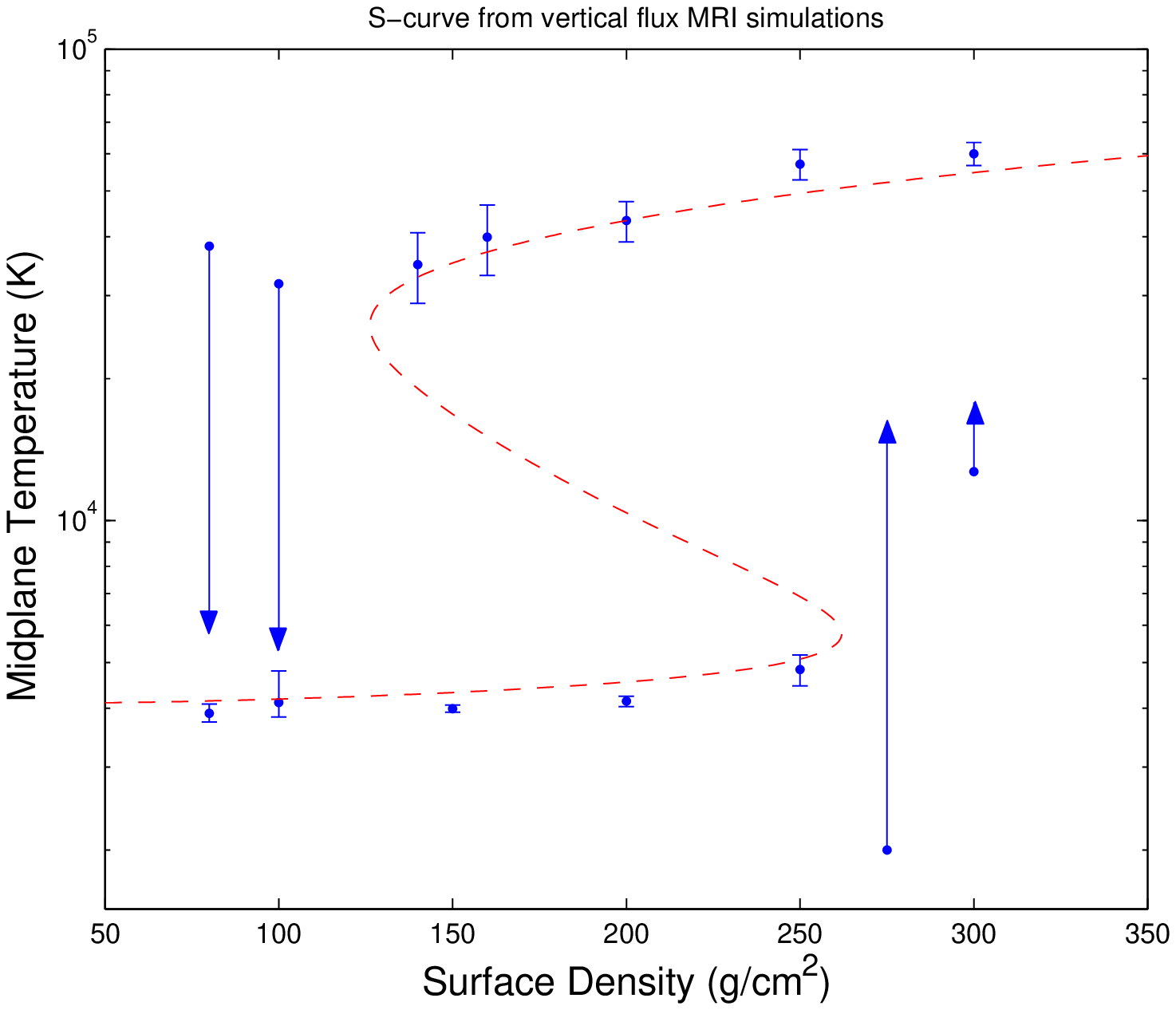}
\includegraphics[width=3.0in,height=3in,angle=0]{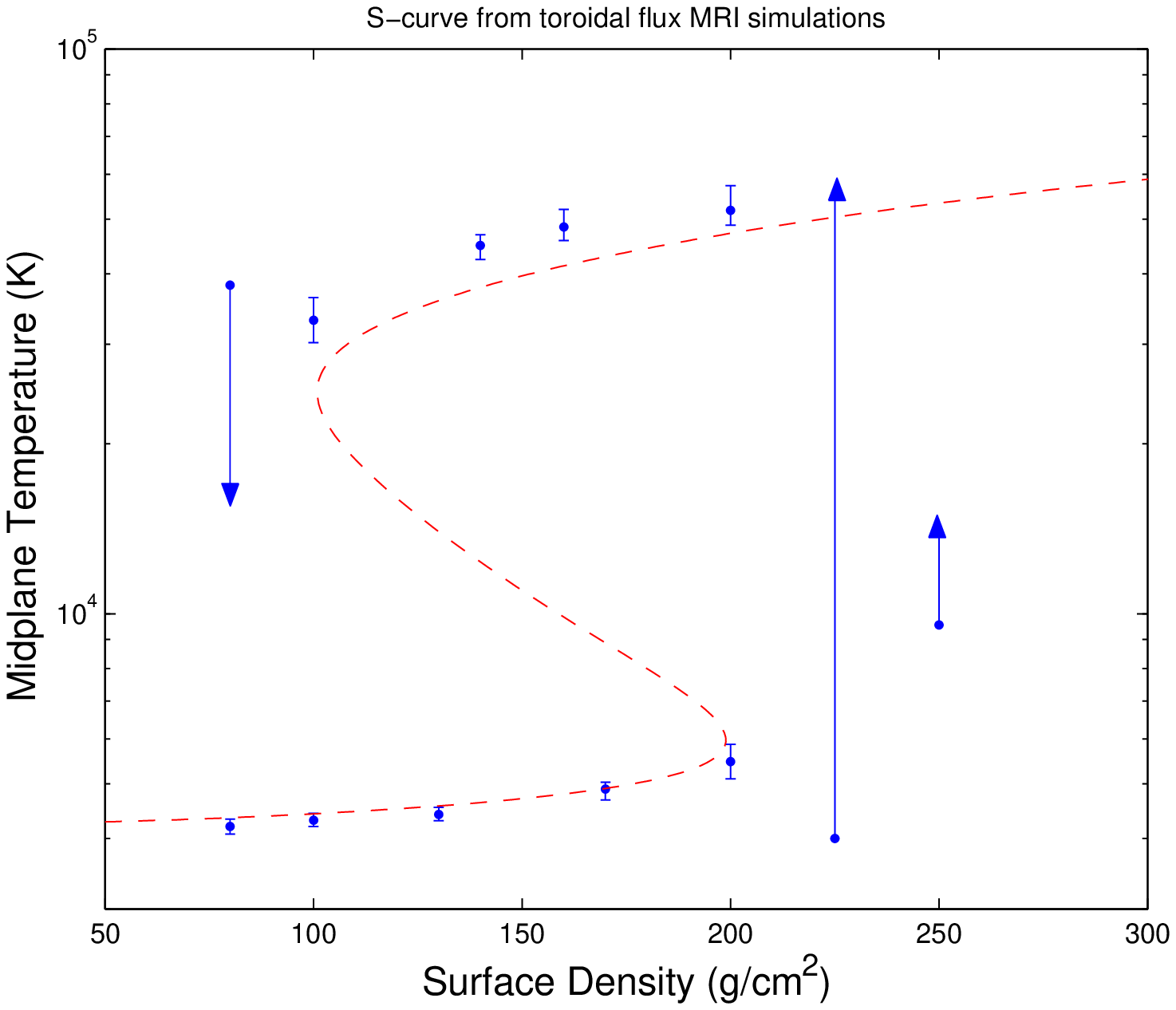}
\caption{ S-curves computed  with NIRVANA and ZEUS
 for the second  parameter set: $(\mu,E,\lambda)= (0.5,5.66,1).$ 
 The first panel corresponds to
 zero net-flux, the second panel to net vertical-flux 
 (initial $\beta=10^4$), and the third panel to
 a net toroidal-flux (initial $\beta=50$).
The dashed curves represent equilibrium S-curves
calculated using the formalism of FLP83, with $\alpha_\text{SS}=
0.009,\, 0.0475,\, 0.07$ for the zero flux, net-vertical flux, and
toroidal flux cases respectively.  Notation is otherwise the same as in
 Fig.~\ref{fig:5}.
 }
\label{fig:6}
\end{figure*}

\begin{table*}
\begin{tabular}{|l|l|l|l|l|l|l|}
\hline
\multicolumn{7}{|c|}{Paramater Set 1: Thermal quasi-equilibrium} \\
\hline
B Configuration & $\Sigma$ & $\langle T \rangle$ & max($T$) & min($T$) &
$\alpha$ & Orbits  \\ \hline\hline
\multirow{5}{*}{Zero Vertical (l.b.)}
&500 & $2.74\times 10^3$ & $2.91\times 10^3$ & $2.57\times 10^3$ &
0.0062 & 240 \\
& 1000 & $2.79\times 10^3$ & $3.42\times 10^3$ &$2.61\times 10^3$ &
$0.0049$ & 170 \\
& 1200 & $3.16\times 10^3$ & $3.51\times 10^3$ & $2.89\times 10^3$ &
$0.0081$ & 350 \\
& 1400 & $3.52\times 10^3$ & $3.99\times 10^3$ & $2.89\times 10^3$ &
0.0076 & 190 \\
& 1500 & $4.10\times 10^3$ & $5.12\times 10^3$ & $2.89\times 10^3$ & 0.0079
& 300 \\
\hline
\multirow{5}{*}{Zero Vertical (u.b.)}
& 600 & $5.92\times 10^4$ & $6.26\times 10^4$ & $5.70\times 10^4$ & 0.0084 &
180 \\
& 800 & $6.97\times 10^4$ & $7.30\times 10^4$  & $6.78\times 10^4$
&0.0083 & 90\\
& 1000 & $8.14\times 10^4$ & $ 8.95\times 10^4$  & $ 7.69\times 10^4$
& 0.0098 & 190\\
&1600 & $10.1\times 10^4$ & $10.5\times 10^4$  & $9.83\times 10^4$ &
0.0082 & 130\\
& 2000 & $11.5\times 10^4$ & $12.7\times 10^4$  & $10.6\times 10^4$ & 0.010
& 170\\
\hline\hline
\multirow{7}{*}{Net Vertical (l.b.)}
& 200 & $2.93\times 10^3$ & $3.05\times 10^3$ & $2.83\times 10^3$  & 0.0348
& 80 \\
& 300 & $3.16\times 10^3$ & $3.35\times 10^3$  & $3.00\times 10^3$ & 0.0342
& 40 \\
&350 & $3.25\times 10^3$  & $3.68\times 10^3$ & $3.04\times 10^3$ &
0.0359 & 90 \\
& 400 & $3.35\times 10^3$ & $3.68\times 10^3$ & $3.17\times 10^3$
&0.0368 & 80 \\
& 500 & $3.71\times 10^3$ & $4.50\times 10^3$ & $3.33\times 10^3$
&0.0387 & 105 \\
& 550 & $4.56\times 10^3$ & $6.27\times 10^3$ & $3.33\times 10^3$ &
0.0384 & 160\\
& 575 & $4.26\times 10^3$ & $5.09\times 10^3$ & $3.33\times 10^3$ &
0.0355 &  100\\
\hline
\multirow{4}{*}{Net Vertical (u.b.)}
& 200 & $3.66\times 10^4$ & $4.11\times 10^4$ & $3.16\times 10^4$ &
0.0362 & 65 \\
& 400 & $5.99\times 10^4$ & $6.87\times 10^4$ & $5.46\times 10^4$
&0.0351 & 140\\
& 650 & $7.95\times 10^4$ & $8.79\times 10^4$ & $7.52\times 10^4$
&0.0375 & 48 \\
& 1000 & $10.1\times 10^4$ & $11.1\times 10^4$ & $9.43\times 10^4$ &
0.038 & 90\\
\hline
\hline
\multirow{3}{*}{Net Toroidal (l.b.)}
& 100 & $3.27\times 10^3$ & $3.44 \times 10^3$ & $3.16\times 10^3$ & 0.0476 &
110 \\
 & 250 & $3.74\times 10^3$ & $4.37\times 10^3$ & $3.46\times 10^3$ & 0.0507 &
 110 \\
 & 500 & $4.01\times 10^3$ & $4.33\times 10^3$ & $3.86\times 10^3$ & 0.0422
 & 75
 \\
\hline
\multirow{5}{*}{Net Toroidal (u.b.)}
 & 150 & $3.72\times 10^4$ &$ 4.12\times 10^4$ & $3.35\times 10^4$ & 0.0442
 & 65
 \\
& 250 &$ 5.36\times 10^4$ &$ 5.82\times 10^4$ &$ 5.04\times 10^4$ & 0.0487 &
118 \\
& 500 & $7.99\times 10^4$ & $8.53\times 10^4$ &$ 7.54\times 10^4$ & 0.0491 & 110
\\
& 750 & $9.97\times 10^4$ & $10.6\times 10^4$ & $9.52\times 10^4$ & 0.0491 &
105 \\
& 1000 &$ 10.3\times 10^4$ & $10.9\times 10^4$ &$ 9.97\times 10^4$ & ... &
65 \\
\hline
\hline
\end{tabular}
\caption{Summary of the inputs and average values of simulations
 undertaken by ZEUS  with parameter  set 1: $\mu=1$, $E=3.5$,
 $\lambda=5$.  The magnetic field configuration is varied as is the
input surface density $\Sigma.$  The initial $\beta$ is $10^4$ for
 the net vertical field runs, and 100 for the net  toroidal field
 runs. Quasi-steady thermal equilibria are attained that
align with one of the two stable branches of the S-curve, with `l.b.' denoting `lower
branch', and `u.b.' denoting `upper branch'.}\label{table1}
\end{table*}

\begin{table*}
\begin{tabular}{|l|l|l|l|l|l|l|l|}
\hline
\multicolumn{7}{|c|}{Parameter Set 2: Thermal quasi-equilibrium} \\
\hline
B Configuration & $\Sigma$ & $\langle T \rangle$ & max($T$) & min($T$) &
$\alpha$ & Orbits  \\ \hline\hline
\multirow{5}{*}{Zero Vertical (l.b.)}
& 100 & $3.51\times 10^3$ & $3.65\times 10^3$ & $3.36\times 10^3$ & 0.0092 &
120 \\
& 300 & $3.74\times 10^3$ & $3.97\times 10^3$ & $3.51\times 10^3$ &
$0.0095$ & 180 \\
&400 & $3.82\times 10^3$ & $4.25\times 10^3$ & $3.50\times 10^3$ &
0.014 & 334 \\
& 500 & $3.95\times 10^3$ & $4.07\times 10^3$ &$3.74\times 10^3$ &
$0.0095$ & 123 \\
& 650 & $4.72\times 10^3$ & $5.35\times 10^3$ & $4.07\times 10^3$ &
0.011 & 197 \\
& 750 & $4.84\times 10^3$ & $5.35\times 10^3$ & $2.55\times 10^3$ & 0.0105
& 176 \\
\hline
\multirow{5}{*}{Zero Vertical (u.b.)}
& 300* & $3.27\times 10^4$ & $4.63\times 10^4$  & $2.04\times 10^4$ & 0.011
& 211\\
& 300** & $4.64\times 10^4$ & $5.24\times 10^4$  & $4.03\times 10^4$ & 0.012
& 185\\
& 400$\dagger$ & $5.52\times 10^4$ & $5.74\times 10^4$  & $4.97\times 10^4$ & 0.0095
& 221\\
&400$\dagger\dagger$ & $5.60\times 10^4$ & $6.12\times 10^4$  & $5.09\times 10^4$ &
0.011 & 226\\
&500 & $6.18\times 10^4$ & $6.37\times 10^4$  & $5.98\times 10^4$ &
0.0095 & 90\\
\hline\hline
\multirow{7}{*}{Net Vertical (l.b.)}
& 80 & $3.90\times 10^3$ & $4.08\times 10^3$  & $3.74\times 10^3$ & 0.036
& 70 \\
& 100 & $4.11\times 10^3$ & $4.80\times 10^3$ & $3.83\times 10^3$  & 0.0425
& 67\\
&150 & $3.99\times 10^3$  & $4.06\times 10^3$ & $3.92\times 10^3$ &
0.038 & 33 \\
& 200 & $4.14\times 10^3$ & $4.24\times 10^3$ & $4.03\times 10^3$
&0.041 & 33 \\
& 250 & $4.83\times 10^3$ & $5.19\times 10^3$ & $4.46\times 10^3$
&0.042 & 35 \\
\hline
\multirow{4}{*}{Net Vertical (u.b.)}
& 140 & $3.49\times 10^4$ & $4.08\times 10^4$ & $2.89\times 10^4$ &
0.045 & 67 \\
& 160 & $3.99\times 10^4$ & $4.67\times 10^4$ & $3.31\times 10^4$
&0.041 & 40\\
& 200 & $4.33\times 10^4$ & $4.75\times 10^4$ & $3.90\times 10^4$
&0.028 & 59 \\
& 250& $5.7\times 10^4$ & $6.12\times 10^4$ & $5.28\times 10^4$ &
0.038 & 25\\
& 300& $6.0\times 10^4$ & $6.34\times 10^4$ & $5.66\times 10^4$ &
0.042 & 27\\
\hline
\hline
\multirow{3}{*}{Net Toroidal (l.b.)}
& 80 & $4.20\times 10^3$ & $4.33 \times 10^3$ & $4.07\times 10^3$ & 0.061 &
38\\
& 100 & $4.31\times 10^3$ & $4.43\times 10^3$ & $4.20\times 10^3$ & 0.068 &
 50\\
 & 130 & $4.41\times 10^3$ & $4.54\times 10^3$ & $4.30\times 10^3$ & 0.071 &
 34\\
 & 170 & $4.89\times 10^3$ & $5.03\times 10^3$ & $4.68\times 10^3$ & 0.08
 & 44
 \\
  & 200 & $5.47\times 10^3$ & $5.87\times 10^3$ & $5.10\times 10^3$ & 0.072
 & 44
 \\
\hline
\multirow{5}{*}{Net Toroidal (u.b.)}
 & 100 & $3.31\times 10^4$ &$ 3.63\times 10^4$ & $3.02\times 10^4$ & 0.085
 & 43
 \\
& 140 &$ 4.49\times 10^4$ &$ 4.69\times 10^4$ &$ 4.24\times 10^4$ & 0.075 &
45 \\
& 160 & $4.84\times 10^4$ & $5.20\times 10^4$ &$ 4.58\times 10^4$ & 0.072 & 37
\\
& 200 & $5.18\times 10^4$ & $5.73\times 10^4$ & $4.88\times 10^4$ & 0.054 &
37\\
\hline
\hline
\end{tabular}
\caption{Summary of the inputs and average values for  simulations
undertaken by NIRVANA adopting  parameter set 2: $\mu=1$, $E=5.66$,
$\lambda=1$. 
The initial $\beta$ is $10^4$ for
the net vertical field runs, and 50 for the net  toroidal field  runs. 
 Some simulations are performed
with the same value of $\Sigma$ but different initial mean
temperatures: * indicates a run begun with $\langle T\rangle =3.85\times 10^{4}$, whereas **
indicates $\langle T\rangle =4.75\times 10^4$; $\dagger$ denotes an initial
$\langle T\rangle=5.20\times 10^4$ and $\dagger\dagger$ denotes $\langle T\rangle =6.00\times 10^4$.
 The larger variations associated with the former pair of cases occurs because the  steady state
thermal equilibrium is expected to be located near the upper bend in the $S$ curve.}\label{table2}
\end{table*}

\subsection{Thermal equilbria: S-curves as viewed 
in the $(\Sigma,\,\langle T\rangle)$ plane}

In this subsection we present the bulk of our simulation results and
plot S-curves for the cases of both parameter sets and for all three
magnetic field configurations. The same qualitative behaviour emerges in
every scenario, which reinforce the robustness of these features.  

Tables \ref{table1} and \ref{table2} summarise the
 simulation results and set-up for the runs that quickly established
 thermal equilibria.
Simulations which underwent a thermal evolution,
corresponding to  either secular  heating or cooling, are described in Tables
 \ref{table3} and \ref{table4}. In Figs \ref{fig:5} and \ref{fig:6} we 
plot S-curves for the
various runs. These show the time-averaged temperature (once thermal
equilibrium is achieved) as a function of surface density $\Sigma$
(a  conserved input parameter). The figures graphically
 summarise some of the data in Tables \ref{table1}-\ref{table4}. 
As in Fig.~4, the blue points with
 error bars 
represent the runs which were begun near a
 stable thermal equilibrium. The arrows represent
`thermally unstable' runs, which were begun at the base of each arrow
and were evolved to a state at the tip of each arrow, at which point
the simulation 
 either approached a stable branch or
was stopped. 

First, the simulations reveal that the volume averaged temperature
fluctuations in the turbulent equilibria are relatively small,
some 10\%, much smaller than for $\dot{M}$. 
This means that the thermal equilibria are in fact very stable:
turbulent fluctuations are not so extreme as to push the system over the
unstable intermediate branch in the $(\Sigma, T)$ diagram. In all the
cases we simulate, equilibrium states never jump branches spontaneously.
 Only at the
`corners' of the S-curve is this a prospect. And on each corner the
system could only move in one direction. As a
consequence, limit cycle behaviour should follow broadly along the lines
predicted by the classical theory.

The `runaway' simulations represent the system switching branches
 on account of the fact that no equilibrium
 state was available.  The
simulations that went from the  vicinity of the lower to upper branches heated up at a
rate $\alpha\Omega$, as expected. 
 The simulations that left the  vicinity  of the upper branch cooled at a varying rate
 determined by $\Lambda$. In some cases turbulence in the latter simulations
 died out once the volume averaged temperature in the box achieved a level much lower
 than the starting temperature $T_0$. This is because the scale height
 in the box becomes so small (of order the dissipation scale) that the
 MRI is suppressed. Similarly, the physical condition of the gas
 becomes mismatched to the box size when the simulations catastrophically
 heat up. Simulations in both cases were discontinued once this
 occurred.

Local simulations that begun near the S-curve
`corners' would often bobble around indeterminately before switching
curves. In these cases the system was near criticality and therefore small
fluctuations in the heating rate change the stability of the system
unceasingly. A sudden change in heating may cause the
local equilibrium to vanish; the gas heats up or cools only for the
heating to readjust and the equilibrium state to reappear. It follows
that in a limit cycle the branch jumping 
 near branch ends can  have a  stochastic component.
 However, we reiterate we do not see  such transitions 
between the \emph{central} regions of the upper and lower branches.

Superimposed upon the numerical S-curves are representative  analytical S-curves
derived from FLP83, in which $\alpha_\text{SS}$ is a user
defined parameter. These $\alpha_\text{SS}$ values were chosen so that good
agreement was obtained with the numerical equilibria on the lower
branch.  Note, in particular, that once we set the analytic curves to
match the lower equilibria, the upper numerical equilibria 
tend to exhibit slightly higher
temperatures than expected from the analytic S-curves. 
Quite generally, the upper numerical branch is better
approximated by alpha models with larger $\alpha_\text{SS}$ values than
required on the lower branches. This could reflect the fact that
the effect of turbulent heating is not accurately described by
the classical $\alpha$ model given the run times of the simulations.
 However, it could also to some extent reflect a bias in the way the
simulations were set up and carried out.

  A close study indicates that the turbulent $\alpha$ weakly depends on the ratio of the average
temperature $\langle T\rangle$ to the reference temperature $T_0$, or
equivalently the mean scale height 
to the reference scale $H_0$. 
Very `hot' (small) boxes with $T> T_0$ exhibit a smaller $\alpha$, while
`colder' (large) boxes yield larger $\alpha$, with relative differences $\sim
30 \%$. This is essentially a numerical effect arising from variations
in the effective Reynolds number and it can operate within any simulation.
 As a consequence, there is some
low level indeterminacy in both
the properties of the equilibrium solutions and the exact locations of
the S-curve corners. Their existence and gross properties, however,
remain. Another, less important, consequence of this `alpha feedback'
is the larger amplitude and intermittent thermal fluctuations in the
saturated turbulent states, as compared to isothermal models.

 We remark that the need for a branch-dependent
$\alpha,$ in the context of $\alpha$ disc modelling,
 was recognised early in
order to obtain
decent matching  of outburst behaviour with observations (Smak 1984a, Lasota 2001).
However,  for the reasons mentioned above, caution
should be exercised in the interpretation of our numerical local $\alpha$ determinations
in this context.
 Finally, note that the levels of turbulence and values of $\alpha$ increase with imposed 
toroidal or vertical magnetic flux. Thus if the flux were to increase as a system transitioned
from a low state to a high state, the higher state would be associated with a larger value of $\alpha.$
 Thus global  disc simulations, for which net-flux need not be conserved
locally, could  potentially lead to such  local $\alpha$ variations.

\subsection{Turbulent stresses and heating}

\begin{figure}
\centering
\includegraphics[width=3in,height=2in,angle=0]{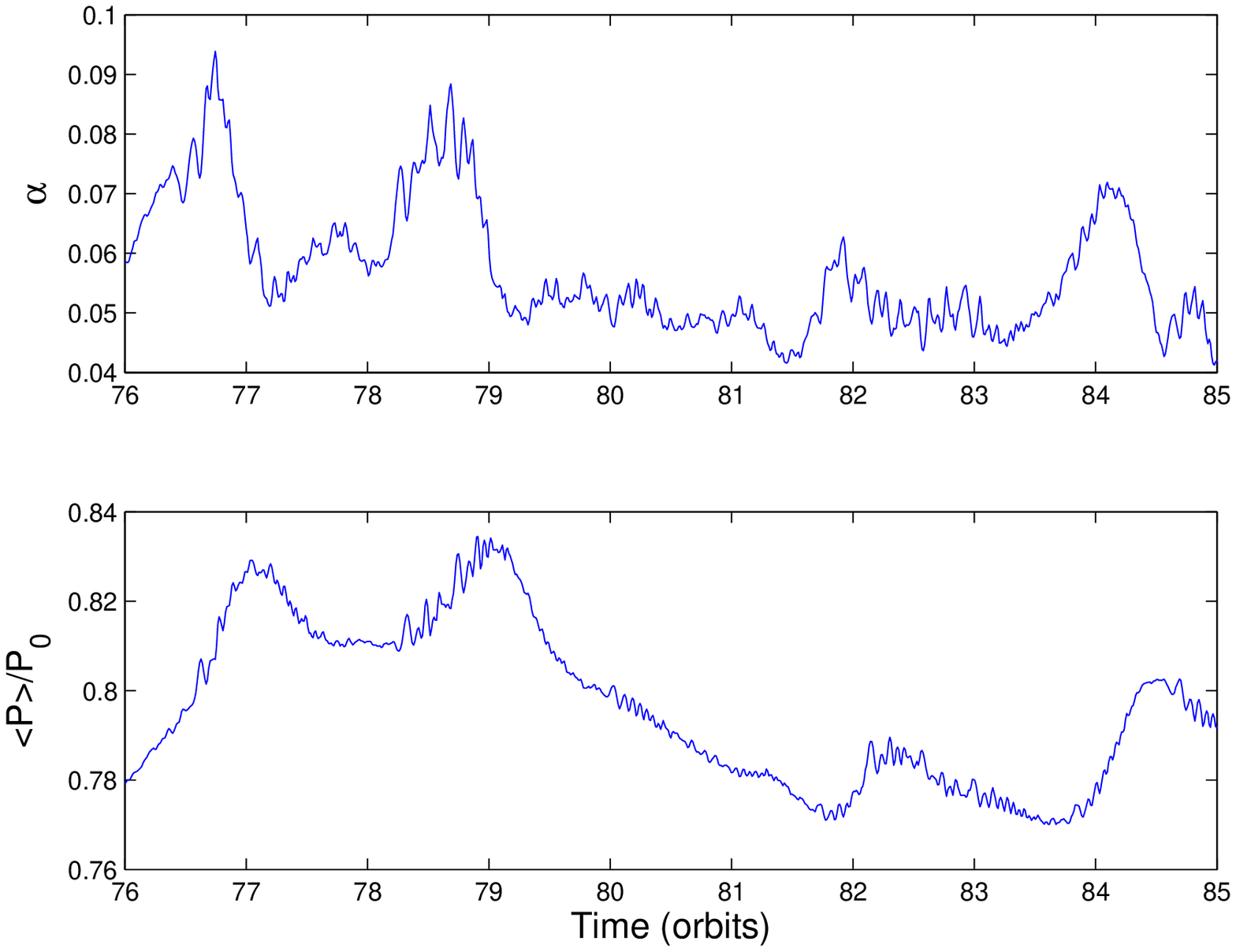}
\caption{Plots of the turbulent stress, quantified by $\alpha$, versus
  time
and the volume averaged pressure versus
  time. Parameters correspond to Set 1; there is a net toroidal-flux associated
  with initial $\beta=100$, and $\Sigma=250$ (UB) (see Table 1).
  It  is clear that  the  latter tracks  the stress imperfectly,
  with much of the short time scale stochastic
   variability  seen in the stress washed out. Gross features are also associated with a
   time-lag of about an orbit. }\label{fig:4}
\end{figure}

We compared the box averaged pressure (and temperature)  with
the turbulent stress in our quasi-steady equilibria. In
Fig.~\ref{fig:4} we plot a  representative evolution of these two
fluctuating quantities. As is clear, on the orbital time-scale there only exists a weak dependence
between pressure and the turbulent stress (described by
$\alpha$). There are three discrepancies. First, pressure \emph{lags behind} the turbulent
stress by roughly an orbit, 
which indicates that on short times the causal relationship is from viscous stress
to pressure and not the other way around, in agreement with Hirose et
al.~(2009). Peaks in $\alpha$ describe intense large-scale events, the
energy of which subsequently tumbles down a cascade to reach  
 dissipative scales. The gas heats up and the pressure
increases. This process, however, is not instantaneous and the
original $\alpha$ signal is significantly modified (`smoothed-out') by
the time that it manifests itself as heating (see Pearson et al.~2004
for similar results in forced isotropic turbulence). Second, 
the volume averaged pressure and $\alpha$ time series  are
dissimilar; in particular, the  volume averaged pressure appears 
at best as a heavily convolved version of
$\alpha$. Lastly, variations
in $\alpha$ are much larger than related variations in the volume averaged pressure,
which are relatively mild, with the former $\sim$30\%, and the
latter less than $5$\%.

This poor correlation between the viscous stress and pressure on short times
casts doubt on the existence of  thermal
 instabilities that rely on these two quantities to be 
functionally dependent (Shakura \& Sunyaev 1976,
Abramowicz et al.~1988). Indeed, numerical simulations of the
MRI in radiation pressure dominated discs reveal no thermal
instability (Hirose et al.~2009). Recently, linear stability
analyses have been conducted 
that allow the pressure to lag behind stress, and vice-versa. 
These
 show that, despite this impediment, instability
can in fact persist, especially with the short time-lags witnessed in our
simulations (Lin et
al.~2011; Ciesielski et al.~2012). It would appear then that the main
obstacle to 
instability in turbulent models
of discs is due to something else, perhaps the stochastic
nature of the viscous stress which is not faithfully 
reproduced in  the behaviour of the volume averaged pressure (Janiuk and Misra 2012).

Finally, note that over \emph{long} times the turbulent stress $T_{xy}$ 
and pressure \emph{are} correlated. Thus turbulent accretion on 
the hot upper branch of equilibria is greater than accretion 
in the colder states (cf. Section 5.1).

\subsection{Numerical convergence}

 Finally, we briefly discuss issues associated with the
numerics. Parameter set 1 was primarily undertaken with ZEUS and
parameter set 2 with NIRVANA, however the two codes were run on an
overlap set of some 8 simulations. The comparison yielded good 
agreement in both $\langle T \rangle$ and
$\alpha$, with discrepancies within the range of the
fluctuations. It follows that not only are our qualitative results
robust with respect to parameter choices and magnetic configurations,
they are robust with respect to numerical method.

In addition, we performed a short study in order to confirm
that the simulations were adequately converged with
respect to resolution and Reynolds number.
A representative sample of runs were taken and we evaluated the
quantities $\alpha$ and $\langle T \rangle$ for different Reynolds
numbers while keeping the Prandtl number constant and equal to 4. 
In summary we found that doubling the kinematic viscosity $\nu$
(halving the Reynolds number) led to
changes in $\alpha$ and $\langle T \rangle$ that were less than
$5\%$. 
This reproduces the very weak Reynolds number dependence uncovered by Fromang (2010).
Moreover, the result held whether we took 64 grid points per $H_0$ or
128 points. Thus we were assured that these quantities were satisfactorily
converged.

Next we kept the diffusion coefficients fixed and varied the
resolution. We evaluated the above quantities at both 64 and 128 grid
cells per $H_0$. The lower resolution runs yielded slightly lower
$\langle T \rangle$ but the discrepancy was within
$10\%$. This small decline is
 understandable because at coarser resolutions 
 more of the turbulent energy is
removed by the grid and thus less captured by physical Ohmic and
viscous heating. As a consequence, there is slightly less direct heating. 
That said, the effect is small and in general no
greater than the statistical fluctuations of the
runs.   It leads  in no way to
any change in the qualitative behaviour exhibited.

\begin{table}
\begin{tabular}{|l|l|l|l|l|}
\hline
\multicolumn{5}{|c|}{Parameter Set 1: Thermal runaways} \\
\hline
B Configuration & $\Sigma$ &  initial $T$ &  final $T$ & Orbits \\ \hline\hline
\multirow{3}{*}{Zero Vertical (c)}
& 300 & $3.7\times 10^4$ &  $6.0\times 10^3$  & 
 60\\
& 400 &  $4.0\times 10^4$ & $3.5\times 10^3$ &
125 \\
& 500 & $4.75\times 10^4$ & $2.7\times 10^3$ &
178 \\
\hline
\multirow{1}{*}{Zero Vertical (h)}
&  1700 & $ 4.5\times 10^3$ & $ 3.4\times 10^4$ & 535\\
\hline\hline
\multirow{3}{*}{Net Vertical (c)}
& 150 & $4.8\times 10^4$  & $6.3\times 10^3$ & 20 \\
& 175 &  $6.5\times 10^4$ & $7.1\times 10^3$ & 31 \\
& 188 &  $6.0\times 10^4$ & $7.8\times 10^3$ & 52 \\
\hline
\multirow{2}{*}{Net Vertical (h)}
& 600 &  $3.3\times 10^3$ & $1.1\times 10^4$ &  101 \\
& 650 &  $3.3\times 10^3$ & $1.4\times 10^4$ &  108 \\
\hline
\hline
\multirow{1}{*}{Net Toroidal (c)}
& 100 &  $5.0 \times 10^4$ & $9.0\times 10^3$ & 14 \\
\hline
\multirow{3}{*}{Net Toroidal (h)}
 & 560  &$3.6\times 10^3$ & $2.5\times 10^4$ & 43  \\
 & 625  &$3.6\times 10^3$ & $4.1\times 10^4$ & 73  \\
 & 750  &$3.0\times 10^3$ & $4.4\times 10^4$ & 68  \\
\hline
\hline
\end{tabular}
\caption{
Summary of the inputs and   range of values found for  simulations
undertaken  ZEUS adopting parameter set 1: $\mu=1$, $E=3.5$,
$\lambda=5$.
 The magnetic field configuration is varied as is the
input surface density $\Sigma$. These simulations resulted in thermal
runaways, i.e. persistent heating
or cooling as indicated by either (h) or (c) .
 This is because they were initially located 
near the corners of $S$ curves in the surface density time averaged mean 
temperature plane. In the latter case this occured on
a time scale of a few orbits in some simulations. In general very large changes to
the mean temperature in these simulations  eventually resulted in the computational box becoming mismatched to the
putative scale height. Accordingly, they were then not continued.
}\label{table3}
\end{table}

\begin{table}
\begin{tabular}{|l|l|l|l|l|}
\hline
\multicolumn{5}{|c|}{Parameter Set 2: Thermal runaways} \\
\hline
B Configuration & $\Sigma$ &  initial $T$ &  final $T$ & Orbits \\ \hline\hline
\multirow{2}{*}{Zero Vertical (c)}
& 100 & $3.18\times 10^4$ &  $3.18\times 10^3$  &
 5.6\\
& 186 &  $3.18\times 10^4$ & $3.19\times 10^3$ &
17 \\
\hline
\multirow{7}{*}{Zero Vertical (h)}
&  686 & $ 8.49\times 10^3$ & $ 1.70\times 10^4$ & 270\\
& 850 &  $6.34\times 10^3$ & $1.78\times 10^4$  & 155\\
& 950 &  $4.75\times 10^3$ & $1.50\times 10^4$  & 54\\
& 1000 & $6.37\times 10^3$ & $4.07\times 10^4$  & 119\\
& 1100 & $2.00\times 10^3$ & $1.17\times 10^4$  & 92\\
& 1200 &  $6.00\times 10^3$& $3.18\times 10^4$ & 254  \\
& 1200 &   $8.00 \times 10^3 $ & $5.92\times 10^4$ & 215   \\
\hline\hline
\multirow{2}{*}{Net Vertical (c)}
& 80 & $3.82\times 10^4$  & $6.37\times 10^3$ & 3.8 \\
& 100 &  $3.18\times 10^4$ & $5.31\times 10^3$ & 5.25 \\
\hline
\multirow{2}{*}{Net Vertical (h)}
& 275 &  $2.00\times 10^3$ & $1.62\times 10^4$ &  81 \\
& 300 &  $1.27\times 10^4$ & $1.78\times 10^4$ &  33 \\
\hline
\hline
\multirow{1}{*}{Net Toroidal (c)}
& 80 &  $3.82 \times 10^4$ & $1.59\times 10^4$ & 5.9 \\
\hline
\multirow{2}{*}{Net Toroidal (h)}
 & 225  &$4.00\times 10^3$ & $5.56\times 10^4$ & 66  \\
 & 250  &$9.55\times 10^3$ & $1.46\times 10^4$ & 43  \\
\hline
\hline
\end{tabular}
\caption{ As in table \ref{table3} but for simulations 
undertaken by NIRVANA and ZEUS adopting parameter set 2: $\mu=0.5$, $E=5.66$,
$\lambda=1$. 
}\label{table4}
\end{table}

\section{Conclusion} 
\label{sec:disc}

In this paper we performed a suite of numerical simulations of
the MRI in local geometry with the ZEUS and NIRVANA codes. Both
viscous and Ohmic heating is included, while
the radiative cooling is approximated by a 
physically motivated cooling function that summarises the strong effect
of temperature
on the ability of the disc gas to retain heat. 
Different magnetic configurations
(zero flux, net-toroidal flux, net-vertical flux) and
parameters were trialed, with little change in the qualitative results.

Our simulations unambiguously exhibit the development of thermal
instability and hysteresis. 
In particular, through a sequence of runs we can sketch out
characteristic S-curves in the phase space of $(\Sigma, T_c)$ and
$(\Sigma, \dot{M})$, which
are central to the classical outburst model. It hence appears that MRI turbulence is
not so intermittent as to endanger the robustness of the cycle.
Temperature fluctuations are well-behaved and relatively small and there
is no spontaneous jumping from one stable branch to the other. 
Only
near the `corners' of the S-curve does significant stochasticity
enter, 
as then the existence or not of a local equilibrium is
uncertain. This feature will add some degree of low level `noise' to
the observed outburst time-series. In addition, the $\alpha$ we record on
the two stable branches differ slightly but systematically. Because of
the constraints of our local model, this result is more suggestive
than anything else. However, it does indicate that in global disc
simulations we may indeed see a systematic difference in the two
branches, as required by the classical theory.

Finally, on the orbital time the turbulent stresses and pressure only weakly depend on
each other in 
our simulations. Pressure always lags behind the turbulent stress, and
thus the causality is from the stress to the pressure, via the
turbulent heating (in agreement with Hirose et al.~2009). Moreover,
the pressure response is a significantly `smoothed out' echo of the
turbulent stress features. Both facts indicate that thermal
instability driven by turbulent heating variations, in which the stress is a
function of pressure, may not operate in real discs.
On long times, however, there is necessarily a feedback of
 the pressure on the stress, which leads to different accretion
 rates on the two branches of the S-curve. 

Our work presents a first step towards
 unifying simulations of full MHD turbulence
with the correct thermal and radiative physics of outbursting
 DNe and LMXBs, and possibly young stellar objects. 
  We have begun with with the most basic model that yields the correct
  physics: local simulations of gas inhabiting a single radius in
  the disc
  with a simple
  radiative prescription. This is a `ground zero' test of the
  compatibility of the MRI with the putative thermal limit cycles of
  outbursting discs. If
  MRI-turbulence had failed in this basic setting it would probably fail
  in more advanced models as well. Now that we are assured of this
  compatability, a variety of further work may be attempted. For instance, simulations
  could be performed in
 vertically stratified shearing boxes with more realistic radiation
 physics (in the flux-limited diffusion approximation with appropriate
 opacities). In addition, cylindrical MRI simulations
 should be performed with the FLP83 cooling prescription, anologous to
 the global $\alpha$ disc calculation of Papaloizou et
 al.~(1983). Such simulations would describe how real turbulence mediates the heating and cooling
 fronts that propagate through the disc during a transition between
 branches.

\section*{acknowledgments}
The authors thank the anonymous referee for his review and Sebastien
Fromang for his comments on an earlier version of the manuscript.
This work was supported by STFC grant ST/G002584/1 
and the  Cambridge high performance computing
service DARWIN cluster.  HNL thanks Tobias Heinemann and
Pierre Lesaffre for coding tips.


\appendix

\section{Radiative cooling model}

In this appendix we detail how the cooling prescription of Section 3.2
was derived. In particular we show how the effective temperature $T_e$
is calculated within each of the three important gas regimes
introduced.

\subsection{Regime 1: the optically thick hot regime}

The radiative cooling function can be written as a flux in
 the diffusion approximation by $\Lambda = \nabla \cdot \bb{F}$ where
\begin{equation}
\bb{ F} =-\frac {4acT^3}{3\kappa \rho}\nabla T ,
\end{equation}
in which $\kappa$ is the opacity, $a =4\sigma/c$ is the Stefan-Boltzmann radiation constant
and $c$ is the speed of light.
Setting the optical depth  as
\begin{equation}\label{tau}
 \tau=\int^{\infty}_z \kappa(x,y,z')\rho(x,y,z') dz',
\end{equation}
  we obtain
the vertical component of $\bb{ F}$ in the form
\begin{equation}
 F_z =\frac {4acT^3}{3}\frac{ \partial T}{\partial \tau} .
\end{equation}
We integrate from near the disc surface where $T=T_e,$ (the effective
temperature),
  and $\tau=\tau_e \sim 1$ to the mid plane where $z=0$, 
$T=T_c$ and $\tau=\tau_c.$
Thus we obtain
\begin{equation}
T^4_c -T_e^4= \int ^{\tau_c}_{\tau_e}\frac{3 F_z}{ac}d\tau \label{thick},
\end{equation}
where $T_c$ is the mid plane temperature.

In the hot optically thick Regime 1, $\tau_c \gg \tau_e\sim 1$
and $T^4_c  \gg T_e^4$ so that (\ref{thick})  implies
\begin{equation}
T^4_c \sim \int ^{\tau_c}_{1}\frac{3 F_z}{ac}d\tau \sim \frac{3\tau_c}{4}T_e^4,
\label{thick1}
\end{equation}
where we have approximated $F_z$ to be constant and equal
to its surface value $acT_e^4/4.$ This equation relates the effective
temperature $T_e$ to the central temperature $T_c$. 

To complete the prescription we
take the scale height to be
\begin{equation}
 H=({\cal {R}}T_c/\mu)^{1/2}\Omega^{-1},
\end{equation}
and set the surface density to
\begin{equation}
\Sigma=2\rho_c H.
\end{equation}
Finally, as most of the optical depth arises from the regions near the midplane,
we approximate Eq.~\eqref{tau} by
\begin{equation} \label{tpres}
 \tau_c=\kappa_c\Sigma/2,
\end{equation}
where $\kappa_c$ is the opacity evaluated at the density $\rho_c$
and the midplane temperature $T_c.$ In this hot ionised regime,
 we can approximate $\kappa$ by the following formula:
\begin{equation}
\kappa = 1.5\times 10^{20}\rho T^{-2.5},
\end{equation}
(FLP83). So for specified $\Omega$,
the above  relationships enable $T_e$ (and hence $\Lambda$) to be related to conditions in
the mid plane, and hence $T_c$ and $\Sigma$.

\subsection{Regime 2: the warm transitional regime}

This regime exhibits a cooler disc midplane that is still well
ionised but surface layers that are $\lesssim 5000$ K, and which are
poorly ionised. As a result, near the photosphere $\kappa$ not only drops
significantly but becomes a steeply increasing function of
temperature. It follows that the requirement $\tau\sim 1$ at the
photosphere  determines the thermal structure of the disc, as in cool
stars (Hayashi  \& Hoshi  1961). Instead of (\ref{thick1}) one must impose
 the condition
$\tau_e \approx 1$, where $\tau_e$ is the optical depth at the location 
 where $T=T_e$. As the optical surface 
is $\sim H $ above the mid plane, it converts to a relationship between the disc surface density
$T_e$ and $T_c$ that leads to the  central unstable part of the 
S-curve when thermal equilibria are considered.

We introduce the quantity $\tau_e^*$ which we define through
$\tau_e^*=\kappa_e\Sigma$ 
with $\kappa_e$ being the opacity evaluated at the \emph{central} density $\rho_c$
but with the photospheric temperature $T_e$. 
 It is anticipated that $\tau_e^*$
 overestimates the optical depth of material
above the disc photosphere $\tau_e$ by some  factor of order unity. This
factor we quantify through the dimensionless constant $E$, so that
\begin{equation} \label{taust}
  E\,\tau_e^* = \tau_e \approx 1 
\end{equation}
Once the dependence of $\kappa_e$  on $T_e$ and $\rho_c$ is specified,
equation \eqref{taust} supplies a means to obtain $T_e$, and hence
$\Lambda$,
when the
disc is in Regime 2. Finally, an approximate functional form of $\kappa$ in this   
 regime of incomplete hydrogen ionisation  is
\begin{equation} \label{kappa2}
\kappa = 10^{-36}\rho^{1/3} T^{10},
\end{equation} 
(FLP83).

\subsection{Regime 3:  the  cool optically thin regime}
In this regime the vertical structure of the disc is isothermal
 with $T\approx T_c$. However, because the disc is optically thin
 in the vertical direction,  $F_z$ is reduced from $acT_c^4/4$ by
a factor $\sim 2\tau_c.$ Thus we set
   \begin{equation} 
    T_e^4 =  2\lambda \tau_c T_c^4,
\end{equation}    
where $\lambda$ is a dimensionless  constant of order unity
that takes account of appropriate frequency averaging and any other necessary corrections
arising from  a more complete discussion of radiation transport.
This condition also connects $T_e$ to $T_c$ and $\Sigma$.
The central optical depth in this regime $\tau_c$ is computed using
Eq.~\eqref{tpres} and
the opacity prescription introduced in Eq.~\eqref{kappa2}.

\subsection{Interpolated form}
The prescription for $T_e$ is given in Eq.~\eqref{Te} for each of the
three regimes. In numerical simulations one could switch from one form to
the other according to the local value of $T_c$; we however employ an
analytic interpolation of the three forms, as in FLP83. In
Eq.~\eqref{Te},
the effective temperature $T_e$ takes one of three forms, which we denote by either $T_1,\, T_2$ or
$T_3$ where the subscript indicates the associated regime. 
These three functions are then interpolated via the
following:
\begin{eqnarray} \label{interp}
T_e^4 &=& \frac{(T_1^4 +T_2^4)T_3^4}{\left[(T_1^4 +T_2^4)^{1/2}+T_3^2\right]^2},\label{cool}
\end{eqnarray}
(FLP83). It is this algebraic expression for $T_e$ that is used in Equation
\eqref{cooling} to evaluate $\Lambda$.
The two free parameters $\lambda$ and $E$
 can be chosen to fit results obtained from
vertical integrations with  standard local $\alpha$ modelling. In
Fig.~(A1) we plot the resulting functional dependence of $T_e$ on the central
temperature $T_c$ with the three regimes indicated on the curve. In
regimes 1 and 3 the effective temperature increases steeply with
central temperature, while in the transitional regime 2 the
dependence is very weak on account of the rapid changes in opacity
near the surface. In this intermediate regime the radiative cooling
rate is very `flat' and, consequently, thermal instability ensues
(cf.\ Eq.~\eqref{stability}).

\begin{figure}
\centering
\includegraphics[width=3.5in,height=2.5in,angle=0]{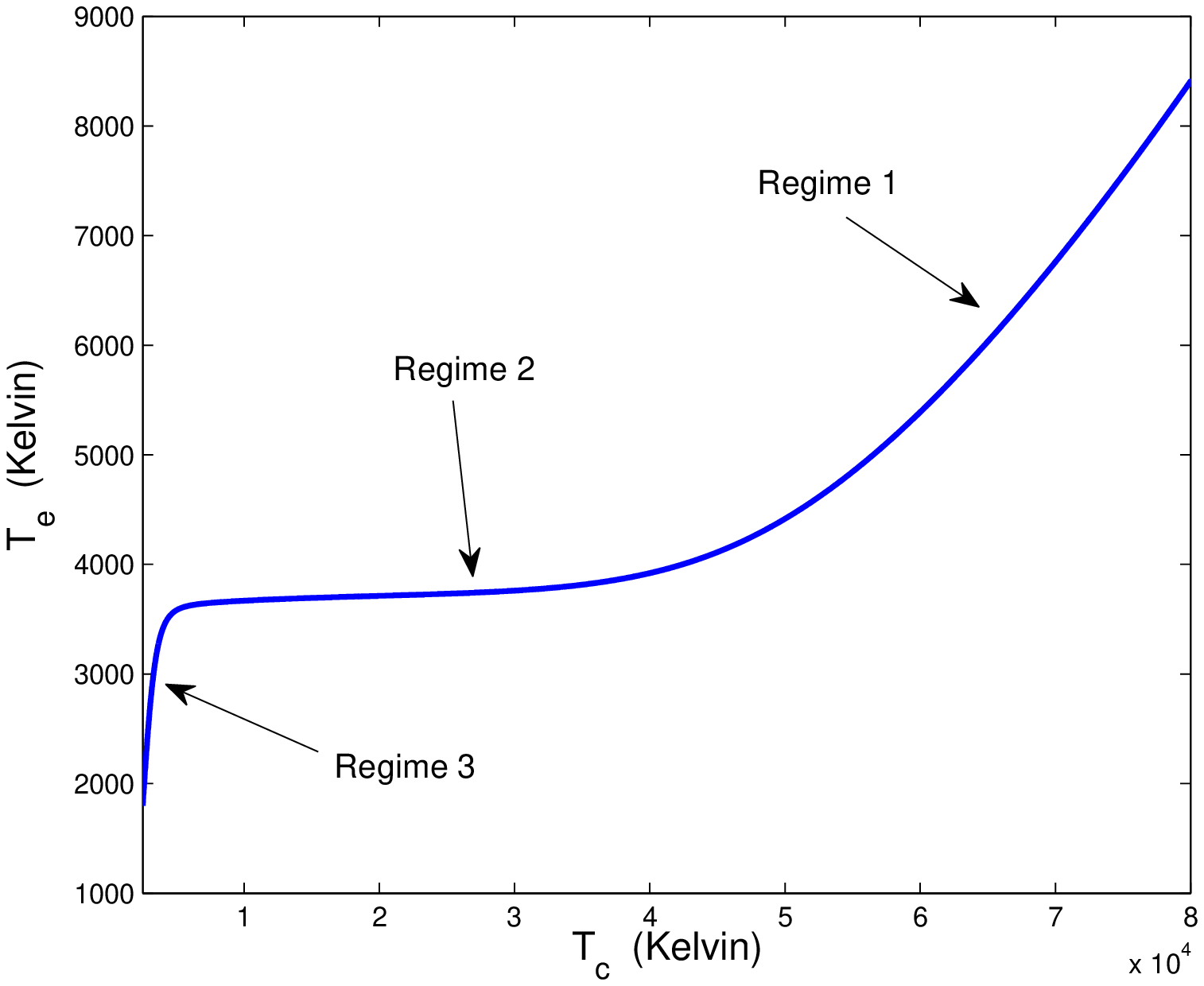}
\caption{The effective temperature $T_e$ as a function of central
  temperature $T_c$ as determined from the FLP83 cooling prescription,
cf.\ Eq.~\eqref{Te}. Parameters are $\mu=1$, $E=3.5$ and $\lambda=5$. In
addition, we have indicated on the curve which regime is dominant.}
\end{figure}

\end{document}